\documentclass[preprint]{aastex631}

\newcommand{\sgra}{\mbox{SGR J1550$-$5418~}}
\newcommand{\sgranosp}{\mbox{SGR J1550$-$5418}}
\newcommand{\sgrb}{\mbox{SGR J1935+2154~}}
\newcommand{\sgrbnosp}{\mbox{SGR J1935+2154}}

\newcommand{\fermi}{\mbox{\textit{Fermi}-GBM~}}

\usepackage{comment, soul}
\usepackage{amsmath}
\usepackage{multirow}
\usepackage{longtable}

\graphicspath{{./}{figures/}}

\shorttitle{Spectro-temporal Investigations of \sgra Bursts}
\shortauthors{Demirer et al.}

\begin{document}

\title{Detailed Time Resolved Spectral and Temporal Investigations of \sgra Bursts Detected with Fermi/Gamma-ray Burst Monitor}

\correspondingauthor{Mustafa Demirer}
\email{mustafa.demirer@sabanciuniv.edu}

\author[0009-0000-9126-7824]{Mustafa Demirer}
\affiliation{Sabanc\i~University, Faculty of Engineering and Natural Sciences, \.Istanbul 34956 Turkey}

\author[0000-0002-5274-6790]{Ersin G\"o\u{g}\"u\c{s}}
\affiliation{Sabanc\i~University, Faculty of Engineering and Natural Sciences, \.Istanbul 34956 Turkey}

\author[0000-0002-1861-5703]{Yuki Kaneko}
\affiliation{Sabanc\i~University, Faculty of Engineering and Natural Sciences, \.Istanbul 34956 Turkey}

\author[0000-0001-9711-4343]{\"Ozge Keskin}
\affiliation{Sabanc\i~University, Faculty of Engineering and Natural Sciences, \.Istanbul 34956 Turkey}

\author[0000-0002-1688-8708]{Sinem Sasmaz}
\affiliation{Department of Physics Engineering, Istanbul Technical University, Istanbul, Turkey}

\author[0000-0002-1688-8708]{Shotaro Yamasaki}
\affiliation{Department of Physics, National Chung Hsing University, 145 Xingda Rd., South Dist., Taichung 40227, Taiwan}

\begin{abstract}

We have conducted a time-resolved spectral analysis of magnetar bursts originating from \sgranosp. Our analysis utilizes a two-step methodology for temporal segmentation of the data. We first generated and fitted overlapping time segments. Subsequently,  we obtained non-overlapping time segments with varying lengths based on their spectral evolution patterns, employing a machine learning algorithm called $k$-means clustering. For the fitting process, we employed three distinct models, namely a modified blackbody (MBB-RCS), a double blackbody (BB+BB), and a power law with an exponential cut-off (COMPT) model. We found that nearly all of the time segments fit well with the COMPT model. Both the average peak energy in the $\nu F_{\nu}$ spectra ($E_{\rm peak}$) and Photon Index parameters follow a Gaussian distribution with the means $\sim$ $30$ keV and $-0.5$, respectively. Furthermore, there is a strong positive correlation between the cooler and hotter temperature parameters of the BB+BB model, and both two parameters show a Gaussian distribution with peaks $\sim$ 4 keV and 12 keV, respectively. Additionally, we found that the distribution of the temperature parameter of the MBB-RCS model can be fitted with a skewed Gaussian function with a peak $\sim$9-10 keV. Lastly, we searched for quasiperiodic spectral oscillations (QPSOs) in the hardness ratio evolution of the bursts. We identified five potential QPSO candidates at frequencies ranging from $\sim$15 Hz to $\sim$68 Hz. We discuss and compare these results with previous studies.

\end{abstract}


\section{Introduction} \label{sec:intro}

Soft Gamma Repeaters (SGRs) are characterized by their emission of intense, short-duration bursts of hard X-rays and soft gamma rays, making them among the most luminous phenomena in these wave bands. These energetic events have attracted high-energy astrophysics communities since they were first discovered in the late 1970s \citep{Mazets_1979}. They were eventually revealed as a distinct class of events; different from any other galactic or extragalactic transient phenomena \citep{Laros__1987, Atteia87}. 

SGRs are now prominent members of a small class of highly magnetized ($B \sim  10^{14}-10^{15}$ G; \citealt{DT92}) isolated neutron stars, which are also known as magnetars. According to the magnetar framework \citep{Thompson_Duncan_1995}, extremely strong internal and external magnetic fields play a pivotal role in burst generation via reconnection or interchange instability in local settings of the neutron star crust. On rare occasions, magnetars also release much longer-duration giant flares releasing extraordinary amount of energies, reaching the level of $10^{45}$ erg \citep{Hurley99, Palmer05}.

\sgra was originally discovered with the Einstein X-ray satellite and received the designation of 1E 1547.0$-$5408 \citep{Lamb_Markert_1981}. The source was identified as magnetar with the detection of radio pulsations with the Parkes Radio Telescope at $P=2.069$ s and the period derivative of $\dot{P}=2.318\times10^{-11}$ s/s; therefore, yielding an inferred dipolar magnetic field strength of $2.2\times10^{14}$ G \citep{Camilo_2007}.

The source did not exhibit any X-ray bursts until 2008, at which point a significant change occurred in its behavior; it entered a series of three burst active episodes. The first episode commenced in 2008 October with several tens of bursts \citep{vonkienlin2012}. The second episode started on 2009 January 22 and became the most burst-active phase of the source to date emitting hundreds of bursts \citep{von12}. This activity episode started with a cluster of bursts with enhanced underlying emission in soft gamma rays, as detected with the Gamma-ray Burst Monitor (GBM) on the \textit{Fermi} Gamma-ray Space Telescope (\textit{Fermi}) \citep{Kaneko09} and also emitted highly energetic bursts with pulsating tails \citep{Mereghetti09}. The third burst active episode of \sgra was in 2009 March-April \citep{von12}. 

Numerous detailed studies have been performed to reveal the spectral properties of short bursts from various magnetars, utilizing both thermal and non-thermal models \citep[see e.g.,][]{Feroci_2004, Israel_2008, Collazzi_2015}. Particularly for the bursts of \sgranosp, \cite{Lin12}, \cite{von12}, and \cite{Kbayrak_2017} conducted time-integrated spectral investigations in various energy ranges. These extensive studies revealed either a combination of thermal models using the two black-body functions (BB+BB) or a power law with an exponential cutoff (COMPT) function successfully described the spectra of most bursts. The COMPT model posits that photons originating from the ignition region undergo successive Compton upscatterings in the magnetosphere due to electron-positron pairs \citep{Lin_2011}.  For this model, \citet{von12} determined an average power-law photon index of $-$0.92, and the average peak energy in the $\nu F_{\nu}$ spectra ($E_{\rm peak}$) of $\sim$40 keV from the time-integrated spectral analysis of 286 bursts of \sgranosp. On the other hand, the BB+BB model proposes the existence of a compact, hotter inner region in the magnetosphere, nested within a larger, colder outer region. This type of structure arises from energy dissipation in the outer regions. Previous time-integrated spectral studies revealed blackbody temperatures of 2--4 keV in the cooler region and 10--15 keV for the hotter emitting region \citep{Feroci_2004, Lin12, von12, Kbayrak_2017}.

Subsequently, \cite{Younes14} performed a detailed time-resolved spectral investigation of bright bursts from \sgranosp. In particular, their primary aim was to address the issue of the brightest time intervals dominating the overall time-integrated burst analysis and achieve a more comprehensive understanding of burst dynamics and spectral properties, by employing  BB+BB and COMPT models. In their COMPT model analysis, a negative correlation between the burst flux and $E_{\rm peak}$ is observed up to a flux limit of $F \approx 10^{-5}$\:erg \:s$^{-1}$\:cm$^{-2}$, beyond which the correlation becomes positive. On the other hand, using the result of the BB+BB model fits, they found that the area of the emission region vs.\,temperature follows a broken power-law of negative trend; For bursts with low flux, the same negative trend is seen across the temperatures while as the burst flux increases a power-law break emerges, which may be attributed to adiabatic cooling \citep{Younes14}. However, their spectral extraction from particular time intervals was done based on a set criterion; In other words, each successive spectrum was accumulated until a certain signal-to-noise ratio was achieved. Although this is a common practice in time-resolved spectral analysis to ensure sufficient statistics, this method of binning could possibly prevent us from seeing the real spectral evolution within the burst, which in turn may hinder the real insight into the spectral properties of magnetars.

In addition to these spectral properties that allow us to probe the emission mechanism of magnetar bursts, observations of quasi-periodic oscillations (QPOs) in the light curves of two galactic giant flares stand out as particularly remarkable features regarding the interiors of magnetars \citep{Israel_2005,Strohmayer_2005,Strohmayer_2006}. It was already predicted by \cite{Duncan_98} that the solid crust of a magnetar and its high magnetic field coupling may result in large-scale magnetic reconfigurations that have the potential to rupture the crust. The X-ray and gamma-ray flux oscillations manifested as QPOs,  could be the result of global seismic vibrations caused by such occurrences. However, due to very short durations of typical magnetar bursts, identifying QPO features in them is more challenging.

For \sgranosp, \citet{Huppenkothen_2014} analyzed 263 short bursts from its 2009 burst storm. By employing a Bayesian method, they identified two QPOs at $\sim$93 Hz, one at 127 Hz, and a broad signal at 260 Hz. They also showed that these frequencies, consistent with those observed in giant flares, are likely linked to global magneto-elastic oscillations \citep{Huppenkothen_2014}. Recently, \cite{Li_2022} investigated a burst from another prolific bursting magnetar, \sgrbnosp, associated with fast radio burst FRB 200428 \citep{Bochenek_2020}. They identified a significant QPO signal at $\sim$40 Hz which marks one of the strongest (3.4$\sigma$) QPO detections in non-giant flare bursts. Similarly, \cite{Xiao_2024} conducted an analysis of QPOs and power density spectra (PDS) from bursts of the  \sgrbnosp. Their study focused on both individual bursts and averaged PDS to identify potential QPO signals. Although no significant QPO detections above 3$\sigma$ were found, the study revealed candidates $\sim$40 Hz in several bursts, consistent with prior findings in the burst associated with FRB 200428. The detected (or candidate) QPOs described above were revealed using time series data; however, 
\citet{Roberts_2023} recently extended this framework to search for quasi-periodic variations in the spectra of magnetar bursts. In particular, they reported the first detection of quasi-periodic spectral oscillations (QPSOs) in the $E_{\rm peak}$ parameter of X-ray bursts from \sgrbnosp, at $\sim$42 Hz. They proposed that these spectral oscillations likely originate from acoustic waves traveling through a magnetized flux tube, which works as an acoustic cavity.
From the observational point of view, it is not known how rare these intriguing cases are as no systematic search for such oscillations in bright magnetar bursts has been performed.

The spectral analysis part of our study was constructed upon the method introduced by \citet{Keskin24}, which investigated the spectral evolution of \sgrb bursts using a novel two-step approach briefly described below \footnote{See \citet{Keskin24} for details of the methodology.}. Clustering-based time-resolved X-ray spectra were fit using the three models: COMPT, BB+BB, and a physically motivated modified blackbody with resonant cyclotron scattering (MBB-RCS). They demonstrated that the COMPT model was preferred for most spectra and revealed correlations between spectral parameters and flux. In this paper, we present the clustering-based time-resolved
spectral analysis of 44 bright bursts from \sgranosp. In particular, instead of generating spectra with uniform time segments or those based on signal-to-noise ratio, we generated spectra for overlapping time segments which were subsequently analyzed as the first phase. Furthermore, to grasp the spectral change points in the bursts, we employed machine learning-based clustering algorithms to create time segments of varying lengths based on the spectral parameters obtained in the first phase. This innovative technique allowed us to gain deeper insights of the burst properties under investigation. In addition, we present the results of our systematic QPSO search with the \sgra bursts where we found five QPSO-burst candidates. This paper is organized as follows: we introduce the instrument, the burst data, and data selection in Section \ref{sec:obs}.  The spectral analysis methods and their results are presented in Section \ref{sec:spectra}. The search for QPSOs and their results are delivered in Section \ref{sec:qpso}. We discuss our both spectral and timing results in Section \ref{sec:dis}.

\section{Observations and Data Selection} \label{sec:obs}

The data used in this study were collected with \textit{Fermi}, which consists of two primary instruments: The Large Area Telescope (LAT) and the Gamma-Ray Burst Monitor (GBM). The LAT is capable of detecting gamma rays with energies in $\sim20$ MeV--$\sim$300 GeV. The GBM has the capability of observing lower energy photons, ranging from $\sim8$ keV to $\sim40$ MeV. The GBM consists of 12 thallium-activated sodium iodide (NaI) scintillation detectors and two bismuth germanate (BGO) scintillation detectors with energy ranges from $\sim8$ keV to $\sim1$ MeV and $\sim200$ keV to $\sim40$ MeV, respectively \citep[For further information about the telescope and the detectors, see][]{Meegan2009, Atwood_2009}. We employed GBM NaI detectors for our analyses, as the majority of emission from typical magnetar bursts occurs below $\sim200$ keV \citep{Lin12, von12, Younes14}. Among the three data types provided by the GBM, we utilized Time-Tagged Event (TTE) data due to its superior temporal (2 $\mu$s) and energy (128 channels) resolution. 

\begin{figure}[!htbp]
    \centering
    \epsscale{1.15}
    \includegraphics[]{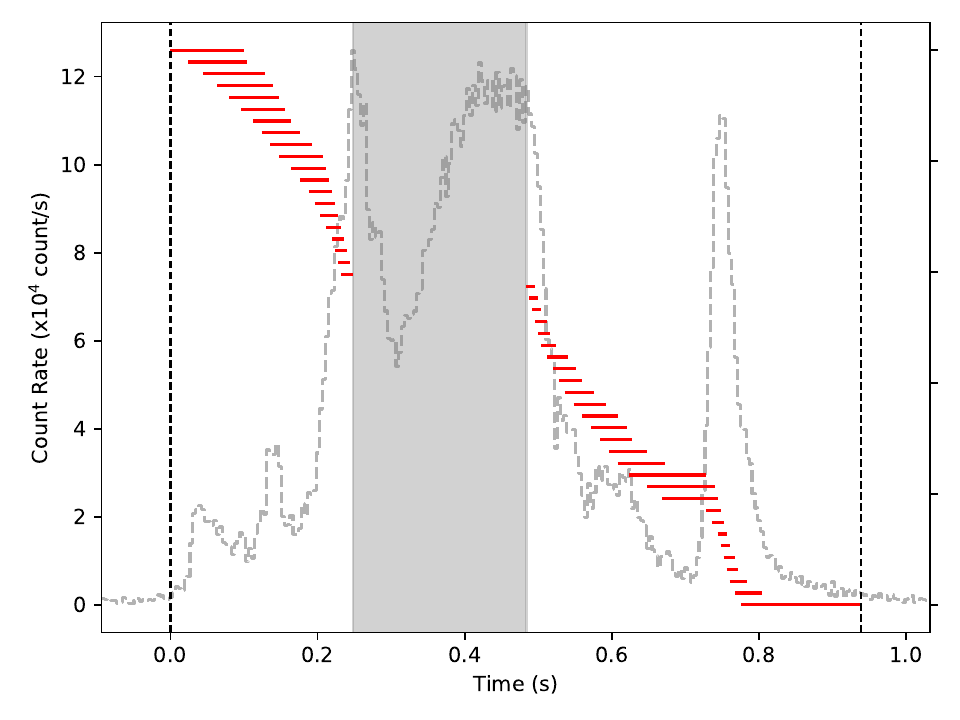}
       \caption{The light curve of an \sgra burst observed on January 22, 2009, at 06:49:48.321 UTC is shown for the brightest detector (n2). Vertical dashed lines indicate the start and end times of the Bayesian Block duration. The red horizontal lines represent 48 overlapping time segments, with each consecutive segment overlapping by 80$\%$. The gray portion corresponds to the saturated part of the burst and spectral analysis is not applied in this range. We note that red lined are intended to display the time span of each segment with an arbitrary y-scale.
       }
    \label{lc_ex}
\end{figure}

We compiled our burst sample from the magnetar burst catalog of \cite{Collazzi_2015}, which includes 386 bursts from \sgra during its 2008-2009 active episodes.  We selected bright bursts with a background-subtracted count above the threshold of 1200 counts (see \citet{Keskin24} the reasoning behind this threshold) in the brightest detector to ensure that the results of our spectral analysis are statistically significant. As a result, we obtained a total of 74 bursts that suit our criterion listed in \autoref{burst_list}.
Among these events, 30 were saturated, namely, the brightest portion of the burst exceeded the limits of the GBM data readout capability of 375,000 counts s$^{-1}$ for all detectors \citep{Meegan2009}. The energy spectra accumulated during these saturated time intervals could suffer from the pulse pile-up effect and may not represent correct energy distribution. Therefore, we excluded the saturated parts of these 30 bursts from our study and subjected only the remaining unsaturated parts to the analysis (see \autoref{lc_ex} for an example). Our analysis was performed in the energy range of 8 to 200 keV, employing 4 ms minimum time resolution, consistent with \citet{Keskin24}. In addition, we excluded the energy range between 30 and 40 keV for the spectral fit statistics calculations to prevent interference from the iodine K-edge\footnote{\url{https://fermi.gsfc.nasa.gov/ssc/data/analysis/GBM_caveats.html}}. 
Nevertheless, we also performed fits with this energy band included, compared their results of two (30$-$40 keV included and excluded) and confirmed that the exclusion does not significantly affect the parameters or their uncertainties.

We used the data collected with three detectors for which the detector zenith-to-source angles are the smallest and less than $60^{\rm o}$ for each burst. In some cases, only two detectors satisfied this angle criterion. Additionally, we excluded detectors whose fields of view were either partially or fully blocked by the spacecraft itself from our analysis\footnote{We checked the blockage using the GBMBLOCK software provided by the GBM team.}. 

%%%%%%%%%%%%%%%%%%%%%%%%%%%%%%%%%%%%%%%%%%%%%%%%%
%%%%%%%%%%%%%%%%%%%%%%%%%%%%%%%%%%%%%%%%%%%%%%%%%
%%%%%%%%%%%%%%%%%%%%%%%%%%%%%%%%%%%%%%%%%%%%%%%%%
%%%%%%%%%%%%%%%%%%%%%%%%%%%%%%%%%%%%%%%%%%%%%%%%%
%EVENT MET: 254299790.321 s

%%%%%%%%%%%%%%%%%%%%%%%%%%%%%%%%%%%%%%%%%%%%%%%%%
%%%%%%%%%%%%%%%%%%%%%%%%%%%%%%%%%%%%%%%%%%%%%%%%%
%%%%%%%%%%%%%%%%%%%%%%%%%%%%%%%%%%%%%%%%%%%%%%%%%
%%%%%%%%%%%%%%%%%%%%%%%%%%%%%%%%%%%%%%%%%%%%%%%%%

\section{Time Resolved Spectral Analysis} \label{sec:spectra}

For the time-resolved spectral analysis, instead of conventional time binning based on the observed signal strength, we applied the method introduced in \cite{Keskin24}: The clustering-based binning approach that identifies significant change points in spectral behavior within the burst. Accordingly, our analysis were done in two phases: I. overlapping time resolved spectral analysis and II. clustered time resolved spectral analysis. To this end, we first segmented our bursts into overlapping time segments within the duration intervals that were calculated using the Bayesian Blocks representation of the burst light curves \citep{scargle2013}. The burst duration was calculated based on data from the brightest NaI detector, with background levels estimated from Bayesian Blocks longer than 4 s. Blocks above this background level were considered burst intervals, and the duration was taken as the time from the first to the last burst block (see Appendix A of \cite{Keskin24} for details). Through this overlapping segmentation, our objective was to capture the spectral evolution as detailed as possible while still keeping sufficient statistics. 

We created the first time segment with a time length that encompasses 1200 burst background-subtracted counts (in the brightest detector). The next time segment began from one-fifth of the duration of the first time segment and continued until the threshold counts of 1200 is accumulated. In doing so, we acquired $80\%$ overlapping time segments. (see \autoref{lc_ex}). As a result, we have obtained a total of 522 overlapping time segments from 74 bursts listed in \autoref{burst_list}, for which spectra in the 8--200\,keV were extracted.  

\subsection{Spectral Analysis and Model Comparison} 

For spectral analysis performed in this study, we used the X-ray Spectral Fitting Package (Xspec; version 12.12.1). We generated Detector Response Matrices for each detector for all untriggered events in our sample with the GBM Response Generator released by the \fermi team. We employed three different spectral models; a modified blackbody with resonant cyclotron scattering \citep[MBB-RCS\footnote{Implemented in Xspec using an additive table model component (atable). The table is available at: \dataset[doi: 10.5281/zenodo.10485159]{\doi{10.5281/zenodo.10485159}}};][]{Lyubarsky_2002, Yamasaki20}{}{}, sum of two blackbodies (BB+BB\footnote{Implemented in Xspec as bbody + bbody.}), and a power law with an exponential cut-off (COMPT\footnote{Implemented in Xspec as a user-defined function: $f(E) = A \exp{[-E(2+\Gamma)/E_{\rm peak}]}(E/50 {\rm keV})^{\Gamma}$.}). The BB+BB and COMPT models are commonly used in magnetar burst analyses as discussed in the introduction. The MBB-RCS model, on the other hand, is physically motivated considering thermal emission—specifically, blackbody radiation modified (MBB; \citealt{lyu03}) by radiative transfer through the trapped fireball interior. This outgoing radiation then undergoes further resonant cyclotron scattering (RCS; \citealt{Yamasaki20}) by magnetospheric particles during magnetar bursts. This model successfully explains typical intermediate bursts from \sgrb by adjusting the effective blackbody temperature $T_{\rm eff}$ of the MBB radiation \citep{Yamasaki20}.
All spectral fits were performed by minimizing the Castor statistics \citep[C-stat,][]{Castor}{}{}. The C-stat we obtain from a fit is based on the maximum likelihood, and it alone does not provide a measure to test the goodness of fit. Therefore, we 
utilized the method suggested by
\cite{Kaastra2017} to calculate the variance and expected value of the C-stat. In turn, we determined acceptable C-stat values (at 3$\sigma$ level) for each photon model of each spectrum. Even though the goodness calculation provides information on the statistical acceptability for each photon model fitting, it does not yield information for comparing different fits with one another. Therefore, we also used the Bayesian Information Criterion \citep[BIC;][]{Schwarz78}{}{} as the model preference metric: 

\[\text{BIC} = -2 \ln \mathcal{L}_{\rm max} + m \ln N = \text{C-stat} + m \ln N. \]

Here, $\mathcal{L}_{\rm max}$ is the maximum likelihood, \textit{m} represents the number of parameters in the spectral model, and \textit{N} describes the number of data points. We compared the difference in BIC values ($\Delta$BIC) for the pairs of continuum model fits (BB+BB vs. COMPT, COMPT vs. MBB-RCS, and BB+BB vs. MBB-RCS. In the comparison of the two models, when the $\Delta$BIC was greater than 10 (corresponding to the Bayes factor of $\sim$150, implying a likelihood ratio confidence level greater than 99\%; \citealt{Bayes95}), we adopted the model with a lower BIC value as the preferred choice. However, in cases where the BIC difference was less than 10, we accepted both models as favorable due to their comparable goodness of fit given that the fit parameters were well constrained.

After fitting all overlapping time-segment spectra with the three models and obtaining their BIC values, we found that 495 spectra (i.e., $\sim95\%$ of the sample) can be modeled well with the COMPT model; namely, the COMPT model fits are either the most preferred based on $\Delta$BIC or have comparable BIC values as that of the alternative model.  The other two, BB+BB and MBB-RCS models, are less preferred and perform similarly; the former model is preferred for $69\%$ of the spectra while the latter model is preferred for $58\%$ of them.

Following the first round of fitting (that is, modeling the spectra of overlapping time segments), we clustered these overlapping time segments in order to form non-overlapping time segments determined by significant changes in spectral properties. For this process, we implemented a machine learning algorithm, called $k$-means clustering \citep{MacQueen1967} from scikit-learn Python library \citep{scikit-learn} to identify significant spectral change points in each of the bursts. The $k$-means algorithm partitions data into $k$ clusters by minimizing intra-cluster variance. Since the COMPT model was preferred by most of the spectra, we employed both the $E_{\rm peak}$ and photon index parameters of the COMPT model along with the midpoint of each time segment as features for the clustering algorithm with appropriate scaling. 
The algorithm requires the number of clusters ($k$) to be specified at initialization. To determine the optimal number of clusters, we developed a method based on the inertia-versus-$k$ graph, using the average inertia of the highest 25\% of $k$ values plus a small offset. The reciprocal of $E_{\rm peak}$ and photon index errors was used as a weight in the inertia calculation to account for parameter uncertainties. The detailed description of $k$-means clustering implementation used in this study can be found in the appendix of \cite{Keskin24}. 

%%%%%%%%%%%%%%%%%%%%%%%%%%%%%%%%%%%%%%%%%%%%%%%%%
%%%%%%%%%%%%%%%%%%%%%%%%%%%%%%%%%%%%%%%%%%%%%%%%%
%%%%%%%%%%%%%%%%%%%%%%%%%%%%%%%%%%%%%%%%%%%%%%%%%
%%%%%%%%%%%%%%%%%%%%%%%%%%%%%%%%%%%%%%%%%%%%%%%%%
%EVENT MET : 254299790.321 s
\begin{figure}[!htbp]
    \centering
    \epsscale{1.15}
    \includegraphics[]{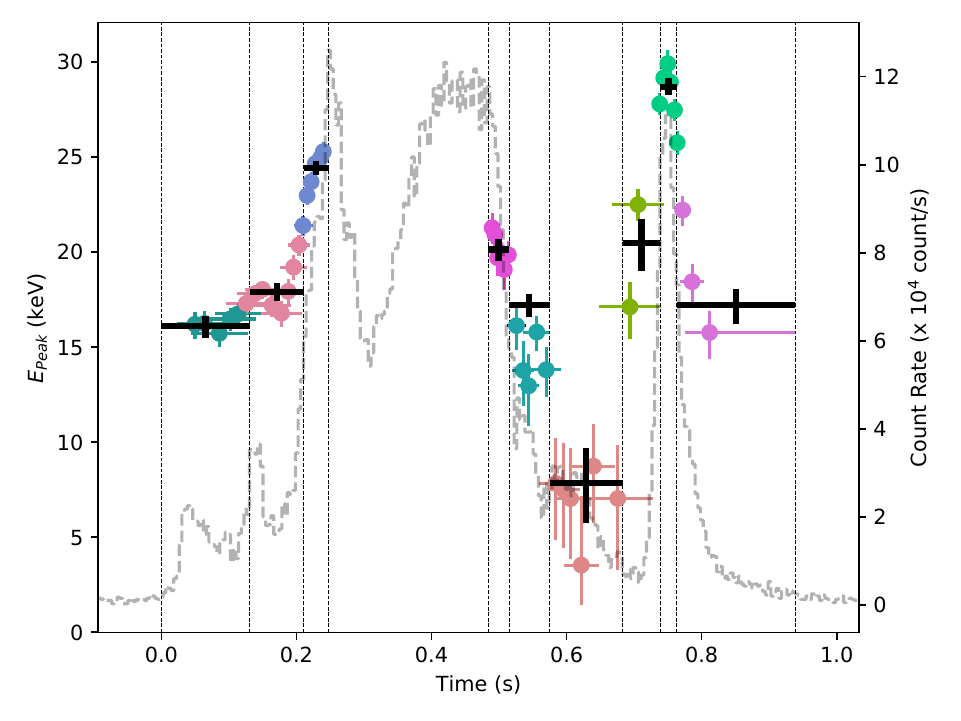}
       \caption{$E_{\rm peak}$ values for 48 overlapping time segments (filled circles with 1$\sigma$ uncertainties) for the same burst shown in \autoref{lc_ex}.  
       The light curve is shown with grey dashed lines (right axis).
       The data points were colored by 9 spectrally-distinctive clusters determined via $k$-means clustering, the intervals of which are shown with vertical dotted lines. 
       Black crosses show the $E_{\rm peak}$ values with 1$\sigma$ uncertainties obtained from the COMPT fits to the nine cluster segments in the second stage of spectral analysis.
       }
    \label{kmeans_plot}
\end{figure}

%%%%%%%%%%%%%%%%%%%%%%%%%%%%%%%%%%%%%%%%%%%%%%%%%
%%%%%%%%%%%%%%%%%%%%%%%%%%%%%%%%%%%%%%%%%%%%%%%%%
%%%%%%%%%%%%%%%%%%%%%%%%%%%%%%%%%%%%%%%%%%%%%%%%%
%%%%%%%%%%%%%%%%%%%%%%%%%%%%%%%%%%%%%%%%%%%%%%%%%

In \autoref{kmeans_plot}, we present an example of clusters obtained with the $k$-means algorithm for one of the bursts in our sample.  As seen in the figure, the time segments that fall between two adjacent clusters identified by the algorithm remain overlapped, and the exact beginning and end times of each cluster need to be systematically determined. To address this issue, we halved the background-subtracted counts in the overlapping regions of the consecutive clusters and assigned half of the counts to the preceding time segment, and the other half to the next time segment. We note that we did not apply the $k$-means algorithm to 16 of the bursts in our sample that provided less than three overlapping time segments, which is insufficient number for clustering. In addition, after applying the $k$-means algorithm to the remaining bursts, we find 14 bursts yielding a single cluster each.
Hence, these 14 events were excluded from further analysis since it was no longer possible to perform time-resolved analysis on these bursts. 

After clustering and above-mentioned exclusion, we obtained non-overlapping time segments ranging from 2 to 9 per burst, with an average of 3 time segments per burst. We then refitted these 151 time non-overlapping segments using our three spectral models. We found that approximately $\sim 93\%$ of them can be fitted with the COMPT model, while approximately $\sim 53\%$ of them can be fitted with the BB+BB model and $\sim 55\%$ with the MBB-RCS model. Note that some time segments being well fitted by multiple models. These results are broadly consistent with the findings from overlapping time segments.

\subsection{Clustering-Based Spectral Analysis Results} \label{sec:res}

\begin{figure}[!htbp]
  \centering
  \includegraphics[width=1\linewidth]{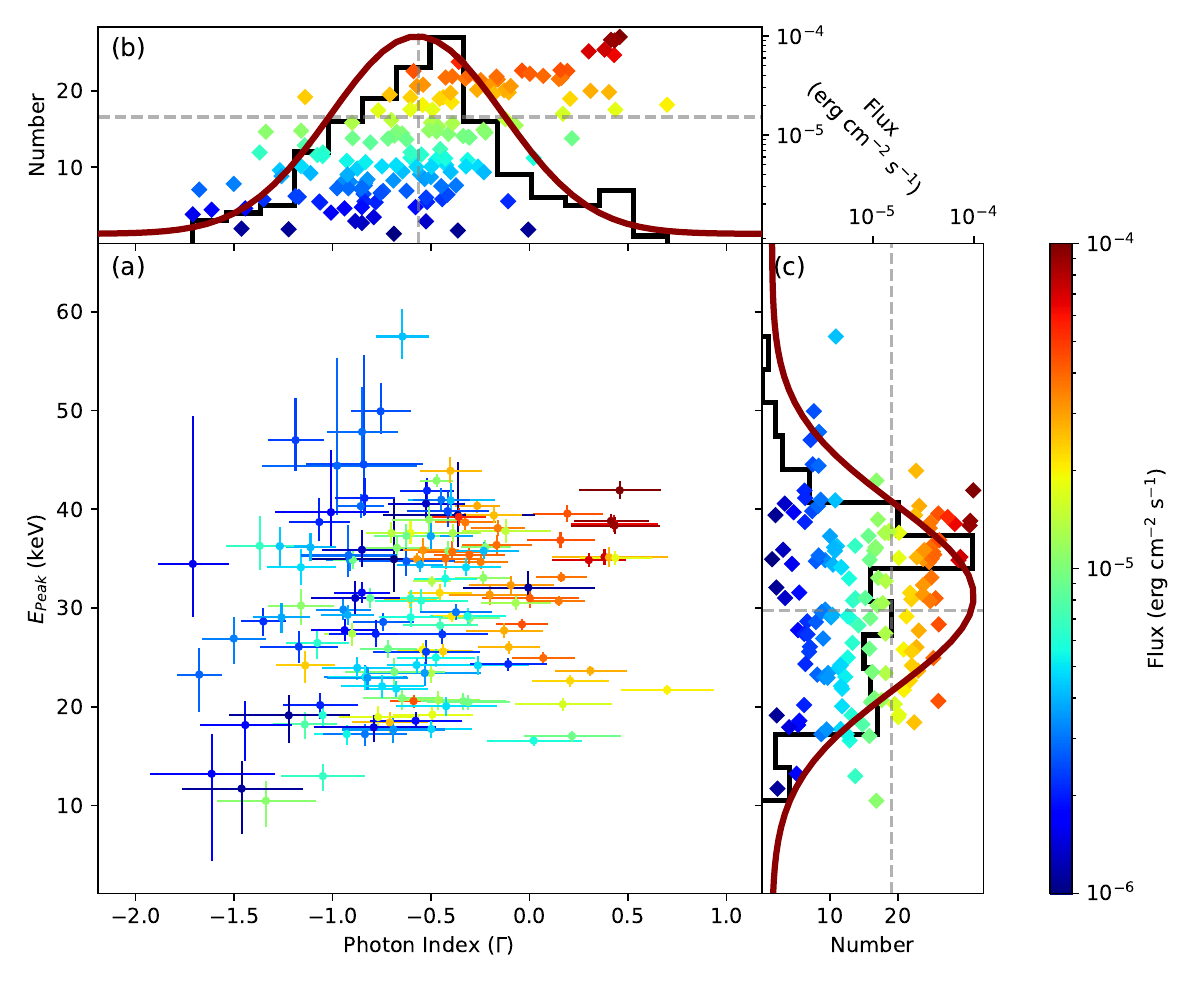}
  \caption{(a) The scatter plot of $E_{\rm peak}$ vs. photon index of the COMPT model fits for 141 spectra. Corresponding energy flux values are color-coded. (b) The distribution of Photon Index values, the best-fit Gaussian function model is shown in brown, and corresponding flux values are shown as diamond data points. The gray dashed line shows the mean value of fluxes. (c) The distribution of $E_{\rm peak}$, the best-fit Gaussian function model is shown in brown, and corresponding flux values are shown as diamond data points. The gray dashed line shows the mean value of fluxes.
  }
  \label{copl_combined}
\end{figure}

Since the COMPT model fits the majority of our spectra, we first discuss the parameters of this model. Note that all uncertainties presented here are at 1$\sigma$ level. We present the scatter plot of the photon index and $E_{\rm peak}$ values of the COMPT model in \autoref{copl_combined} along with the distributions, all color coded by the energy flux. We find that the photon indices ($\Gamma$), which range from --2 to 1, are distributed as a Gaussian (\autoref{copl_combined}b) with the mean value of $-$0.57$\pm$0.04 and the width
of 0.43$\pm$0.04 (reduced chi-square, $\chi^2_\nu$ = $0.83$). The distribution of $E_{\rm peak}$ also follows a Gaussian (see \autoref{copl_combined}c) whose mean is at 30.65$\pm$0.95 keV and with a width of 8.46$\pm$0.98 keV ($\chi^2_\nu$ = $0.86$). The energy flux of these spectra in the 8$-$200 keV energy range from 1$\times$10$^{-6}$ to 9.82$\times$10$^{-5}$ erg cm$^{-2}$ s$^{-1}$, while the energy fluence values are between 1.9$\times$10$^{-7}$ and 1.6$\times$10$^{-6}$ erg cm$^{-2}$.

We find no correlation between the parameters of the COMPT model (i.e., $E_{\rm peak}$ and $\Gamma$). On the other hand, we find a positive correlation between the index and the corresponding flux (Spearman's rank order correlation coefficient, $\rho$ = 0.62 and the chance probability of such correlation to occur from a random data set, \textit{P} = 9.8$\times10^{-18}$).
As seen in \autoref{copl_combined}, the spectra with the highest flux yield photon indices around 0.5 and $E_{\rm peak}$ tightly clustered near 40 keV. Meanwhile, $E_{\rm peak}$ of the spectra at lower flux levels span nearly in the same range from about 15 keV to 45 keV. 

\begin{figure}[!htbp]
  \centering
  \includegraphics[width=1\linewidth]{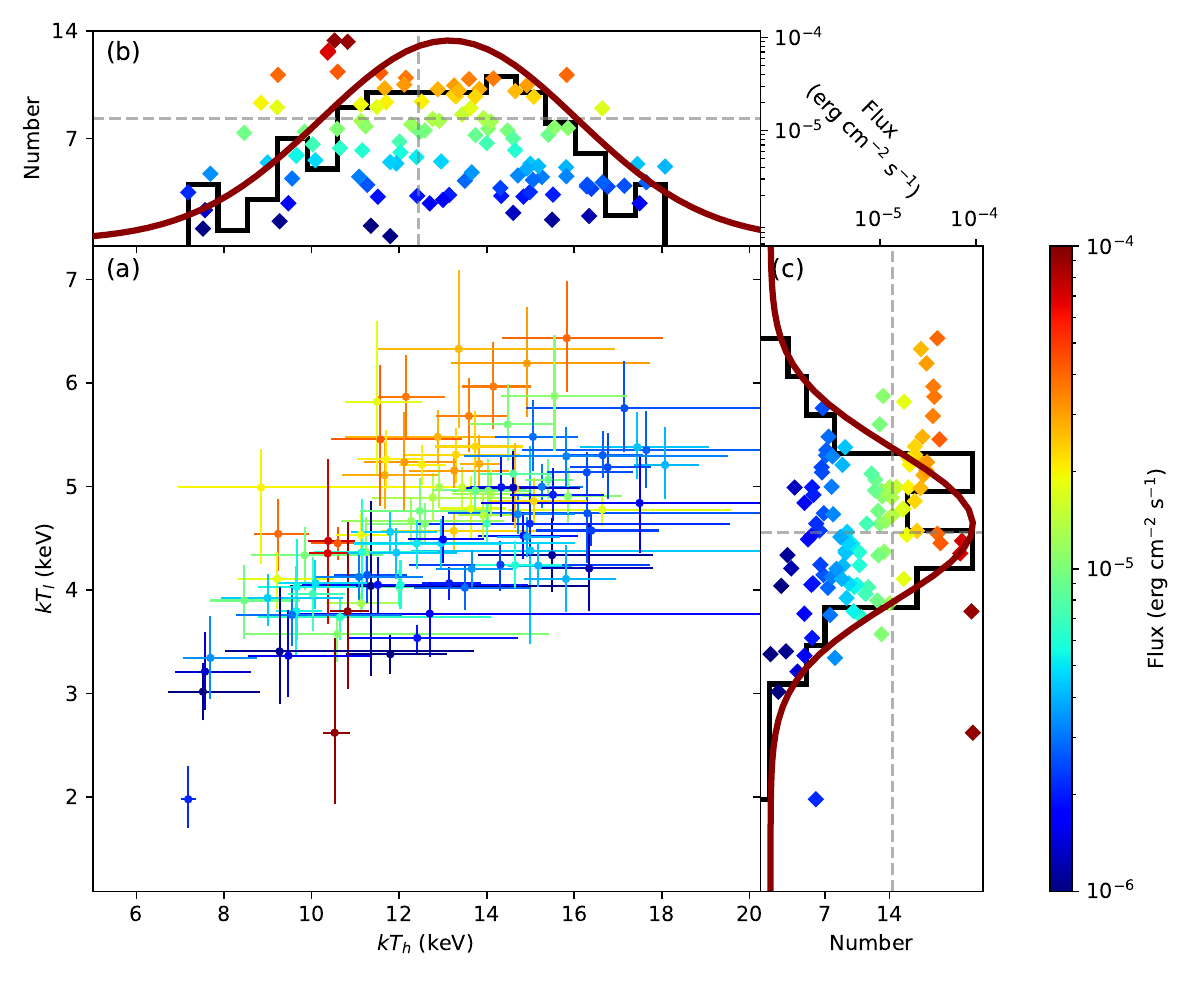}
  \caption{(a) The scatter plot of $kT_{ l}$ vs $kT_{ h}$ parameters that can be described with BB+BB (80 spectra). Corresponding flux values are color-coded. (b) The distribution of $kT_{ h}$, the best-fit Gaussian function model is shown in red, and corresponding flux values are shown as diamond data points. The gray dashed line shows the mean value of fluxes. (c) The distribution of $kT_{ l}$ values, the best-fit Gaussian function model is shown in red, and corresponding flux values are shown as diamond data points. The gray dashed line shows the mean value of fluxes. 
  }
  \label{bb_bb_combined}
\end{figure}

Regarding the thermal models, we present the scatter plot of the two $kT$ values of the BB+BB model in \autoref{bb_bb_combined} along with the distributions, again color-coded by the energy flux. We find that the cooler component ($kT_l$) of the BB+BB model exhibits a Gaussian distribution which peaks at 4.64$\pm$0.07 keV with a width of 0.72$\pm$0.07 ($\chi^2_\nu$ = $0.83$), and it is characterized by a narrow spread between 2 to 7 keV (\autoref{bb_bb_combined}c). However, the hotter component ($kT_h$) of the same model exhibits a much broader distribution, spanning from 6 to 20 keV, and its Gaussian fit curve value has a peak at 13.16$\pm$0.22 keV and a width of 2.93$\pm$0.26 ($\chi^2_\nu$ = $0.90$) (\autoref{bb_bb_combined}b). We observe a statistically significant positive correlation between the parameters, $kT_{ l}$ and $kT_{ h}$, with $\rho = 0.59$ and a $P$-value of $1.3 \times 10^{-11}$. Additionally, we find a moderate but significant positive correlation between $kT_{l}$ and the energy flux ($\rho = 0.41$, $P = 1.1 \times 10^{-5}$), suggesting that higher flux levels are typically associated with higher values of $kT_{l}$. Conversely, $kT_{h}$ does not exhibit any significant correlation with the energy flux. For intermediate flux levels, $kT_{l}$ values are mostly confined to the range of 3--6 keV. In contrast, $kT_{h}$ values span nearly the entire range from 7.5 keV to $\sim$20 keV across all flux levels, except for the highest flux segments, where $kT_{h}$ is located in a narrower range between 10 and 11 keV.

\begin{figure}[!htbp]
  \centering
  \includegraphics[width=0.75\linewidth]{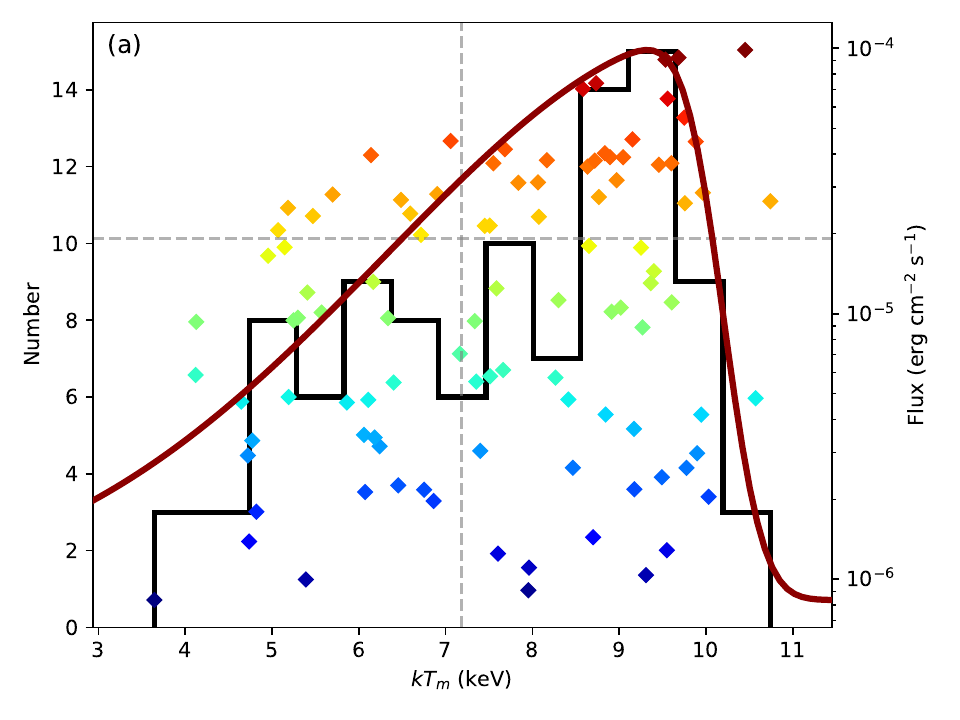}
  \caption{Time segment distribution of the MBB-RCS model temperature, $kT_m$ for 83 spectra. Flux values of each individual time segment are shown in logarithmic scale and color-coded diamond data points. The best skewed Gaussian fit is drawn in red. The gray dashed lines shows the mean value of fluxes and $kT_m$ parameter respectively.}
  \label{mbb_rcs_combined}
\end{figure}

Finally, the MBB-RCS model temperatures ($kT_{ m}$) are presented in \autoref{mbb_rcs_combined}.  The distribution of $kT_{ m}$ is between $kT_{ l}$ and $kT_{ h}$ values of the BB+BB model as expected, ranging from 4 to 11 keV (\autoref{mbb_rcs_combined}a). Even though the sample size is limited, the distribution of $kT_{ m}$ values is more complex than a Gaussian-like distribution, being consistent with a skewed Gaussian distribution. 
The skewed Gaussian fit has a mean value of 10.20$\pm$0.18 and a width of 3.89$\pm$0.65 ($\chi^2_\nu$ = $0.70$). Regarding the flux dependence of the MBB-RCS model, lower and middle flux values are distributed over all $kT_m$ values. However, the highest flux values are accumulated around the 8-11 keV region.

\section{Search for Quasiperiodic Spectral Oscillations} \label{sec:qpso}

In addition to spectral investigation, we extended our analysis to explore any presence of quasiperiodic oscillations in \sgra bursts. In this effort, we were motivated by the findings of \cite{Roberts_2023}, which report possible QPSO signatures in \sgrbnosp. \sgra is another prolific magnetar and numerous bursts are bright enough, well suited for such a search. However, \sgra bursts in our sample consist of very few time segments. Therefore, it was not possible to utilize the $E_{\rm peak}$ parameter to search for potential QPSOs. In order to extend the systematic search for QPSO in a larger burst sample, we performed temporal search using the hardness ratio (HR) instead. This can be justified given the fact that the photon index parameter remains nearly constant during all of the bursts. 
(HR variations closely follow the time evolution in $E_{\rm peak}$). Moreover, we diagnosed the use of HRs for QPSOs using the GBM data of \sgrb bursts that shows prominent QPSO reported by \citet{Roberts_2023}. We find the HR oscillations at around 45 Hz consistent with the reported value with a chance occurrence probability of 0.0068.

For HR evolution, we first obtained a light curve with 4 ms time resolution for each burst by combining the data collected with the detectors included in the spectral analysis. We then calculated HRs in the 8--200 keV energy range with varying energy pivots ($E_{\rm piv}$) between 10--50 keV, where $E_{\rm piv}$ represents the energy threshold separating the soft and hard energy bands. We defined HR as the ratio of background subtracted counts in two different energy bands: $E_{\rm piv}-$200 keV and 8$-E_{\rm piv}$. In order to avoid negative background-subtracted counts of the light curve and to obtain statistically reliable HR evolution throughout each burst, we required that an HR must be constrained within at least a 3$\sigma$ level ($\ge$3 times of its own error). Therefore, if the background-subtracted counts in a 4 ms time bin do not satisfy this criterion, we added the counts of the next time bin until a reliable HR was obtained.

In the QPSO search process, we first eliminated the underlying long-term variations in the HR curve by fitting a third-degree polynomial to the HR data and analyzed the residuals. We then employed two different methods for our analysis: the Lomb-Scargle Periodogram (LSP; \citealt{Lomb_1976, Scargle_1982}) and Weighted Wavelet Z-transform (WWZ; \citealt{Foster_1996}). The LSP was chosen because it is a well-suited tool for time series analysis of unevenly sampled datasets. However, since the LSP does not account for temporal variations, we complemented it with the WWZ, which provides time-resolved spectral analysis. For the WWZ method, we employed a Fortran code, as provided by \cite{Foster_1996}, and for the LSP method we utilized the Astropy package \citep{Astropy_2022}. 

The frequency range used in both methods was obtained as follows: The minimum frequency was set to 10 Hz, based on the durations of typical magnetar bursts ($\sim$0.1 s). The maximum search frequency was obtained by taking the minimum spacing of HR steps (i.e., $\Delta$t = 2 ms) into account, as 1/(2$\Delta$t) = 250 Hz. This is analogous to the Nyquist frequency in searches of evenly sampled time series data. 

In LSP, we used 20 samples per peak corresponding to a frequency resolution of 0.1 Hz and quantified the significance of detected peaks using the False Alarm Probability (FAP) calculated using the method of \cite{Baluev_2008}. For the WWZ method, we adopted the same frequency resolution. We tested various Gaussian window sizes (c) to examine their impact on the QPSO results but found no noticeable differences. Therefore, we selected $c=0.005$, as this value is an order of magnitude smaller than the shortest QPSO candidate duration. In addition, we utilized F-statistics values to determine QPSO significance. 

\begin{figure}[!htbp]
    \centering
    \epsscale{1.55}
    \includegraphics[width=0.9\linewidth, trim = {95 40 75 50}, clip]{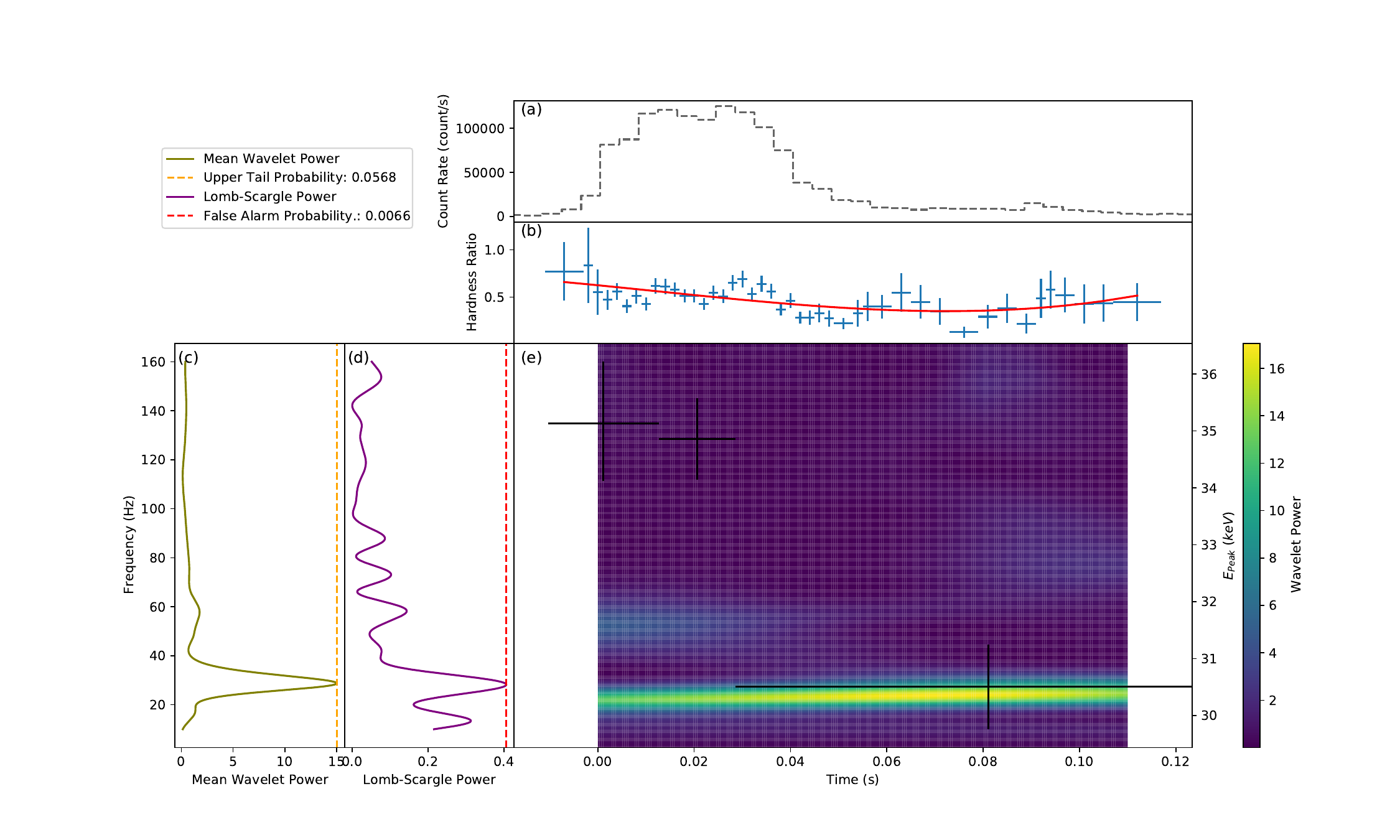}
    \includegraphics[width=0.9\linewidth, trim = {95 40 75 50}, clip]{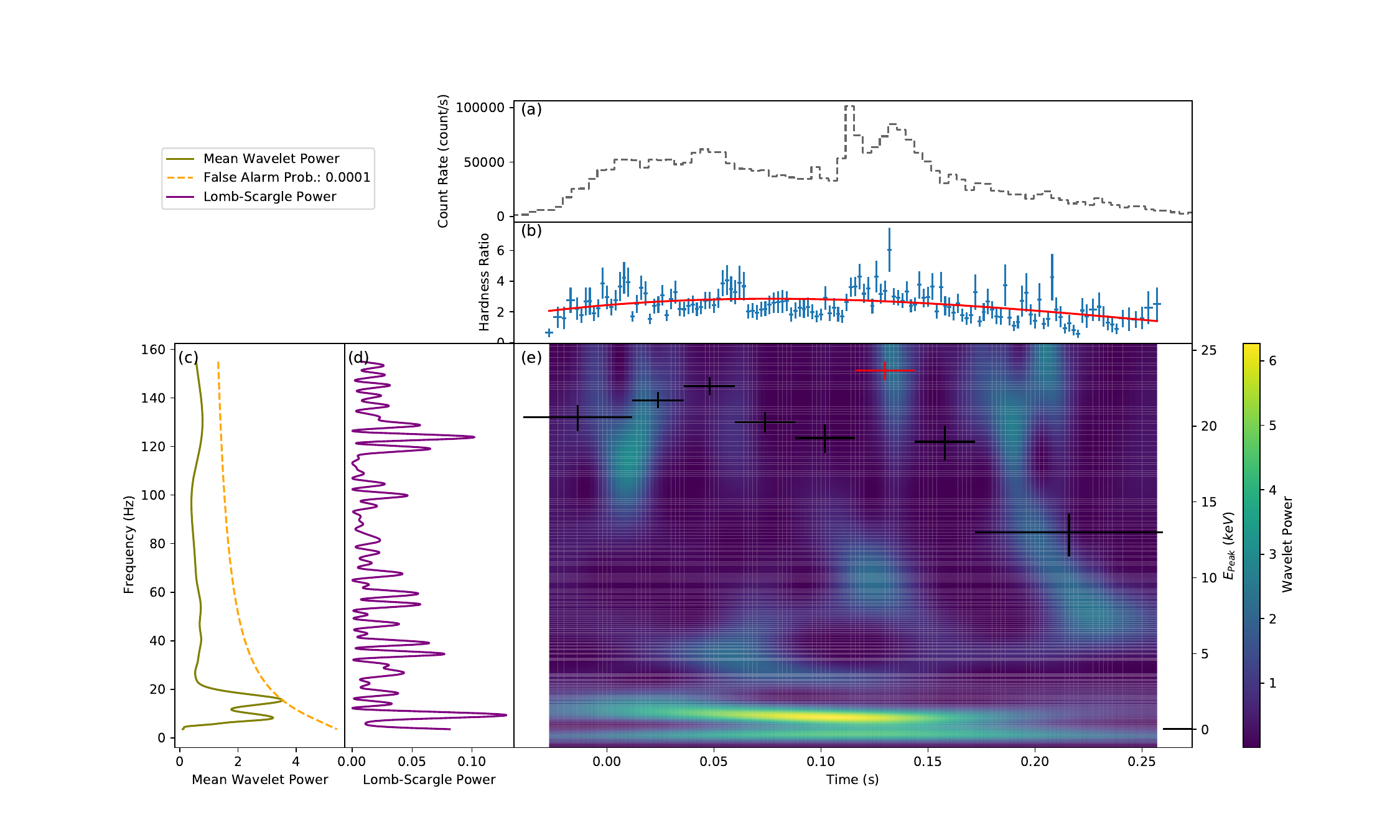}
       \caption{(top panel) Combined plot of most significant candidate QPSOa (MET: 254366383.448) and (bottom panel) Combined plot of candidate QPSOb (MET: 254299756.841). In both figures, (a) the light curve (b) the hardness ratio versus time, fitted with a third-degree polynomial (c) the time-averaged WWZ plot and (d) the Lomb-Scargle periodogram as a function of frequency, are shown. In panels (c) and (d), the maximum values are indicated by dashed lines, with their corresponding $p$-values provided in the legend. Please note that p-value cannot be calculated for the LSP method in QPSOb. (e) the WWZ contour plot with time and frequency axes, where power is color-coded. Overlaid on panel (e) are the $E_{\rm peak}$ values for each time segment derived from the spectral analysis. All time segments, except for the sixth time segment of QPSOb (highlighted in red), are best described by the COMPT model based on BIC comparisons. Although the analysis spans the 10–250 Hz frequency range, the plots are limited to 160 Hz for clarity.
       }
    \label{qpo_plot1}
\end{figure}

For the WWZ method, we first searched the entire burst interval. We identified the time interval(s) in the wavelet periodograms corresponding to a $p$-value of less than approximately 0.1, and further narrowed down the precise search interval of the QPSO candidate using both methods (see color plot in \autoref{qpo_plot1}).

The procedure described above was repeated for all bursts and across multiple pivot point energies ranging from 15 keV to 30 keV in 1 keV increments. We systematically examined each burst with all $E_{\rm piv}$ values and identified 5 bursts with QPSO candidates ranging from 15.2 Hz to 67.84 Hz, each detected at a specific fixed $E_{\rm piv}$ value. The candidate signals in the resulting power spectra were fitted with a Lorentzian function to determine their peak frequency values, and the full width at half maximum (FWHM). Additionally, coherence values (Q = Peak Frequency / FWHM) were calculated to assess the broadness of the peaks. All resulting Q values exceed 2, which is the conventional threshold to call a feature as a QPO \citep{vanderklis_2006}. These results along with $p$-values (chance probability) are tabulated in \autoref{tab:qpsos}.

\begin{deluxetable}{cccccccccccc}[!htbp]
%\rotate

\tabletypesize{\scriptsize}

\tablecaption{The QPSO candidates using WWZ and LSP methods.} 

\tablehead{
  \colhead{} & \colhead{} & \colhead{Begin} & \colhead{End} & \colhead{} & \colhead{Initial} & \colhead{Final} & \multicolumn{4}{c}{WWZ (top), LSP (bottom)} \\
  \cline{8-12}
  \colhead{Name} & \colhead{MET} & \colhead{Time} & \colhead{Time} & \colhead{$E_{\rm piv}$*} & \colhead{$E_{\rm peak}$} & \colhead{$E_{\rm peak}$} & \colhead{Peak Frequency} & \colhead{HWHM} & \colhead{Coherence (Q)} & \colhead{Cycle} & \colhead{$p$-value}  \\
  \colhead{} & \colhead{} & \colhead{(s)} & \colhead{(s)} & \colhead{(keV)} & \colhead{(keV)} & \colhead{(keV)} & \colhead{(Hz)} & \colhead{(Hz)} & \colhead{} & \colhead{Count} & \colhead{} 
}
\startdata
\multirow{2}{*}{QPSOa} & \multirow{2}{*}{254366383.448} & \multirow{2}{*}{-0.01$^{\dagger}$} & \multirow{2}{*}{0.11$^{\dagger}$} & \multirow{2}{*}{29} & \multirow{2}{*}{$35.14^{+1.08}_{-1.02}$} & \multirow{2}{*}{$30.52^{+0.75}_{-0.73}$} & 28.97(8) & 3.55(4) & 4.09 & 3.48 & 0.0568 \\
 & & & & & & & 27.90(1) & 5.43(5) & 2.57 & 3.35 & 0.0066 \\
 \hline
\multirow{2}{*}{QPSOb} & \multirow{2}{*}{254299756.841} & \multirow{2}{*}{-0.03} & \multirow{2}{*}{0.26} & \multirow{2}{*}{15} & \multirow{2}{*}{$20.58^{+0.82}_{-0.90}$} & \multirow{2}{*}{$12.98^{+1.27}_{-1.55}$} & 15.73(4) & 1.33(5) & 5.89 & 4.56 & 0.0001 \\
 & & & & & & & --- & --- & --- & --- & --- \\
 \hline
\multirow{2}{*}{QPSOc} & \multirow{2}{*}{254291434.732} & \multirow{2}{*}{0.0$^{\dagger}$} & \multirow{2}{*}{0.15$^{\dagger}$} & \multirow{2}{*}{15} & \multirow{2}{*}{$40.32^{+1.25}_{-1.15}$} & \multirow{2}{*}{$40.32^{+1.25}_{-1.15}$} & 27.79(4) & 4.02(1) & 3.46 & 4.17 & 0.1021 \\
& & & & & & & 29.45(2) & 4.07(7) & 3.62 & 4.42 & 0.012 \\
\hline
\multirow{2}{*}{QPSOd} & \multirow{2}{*}{254302590.741} & \multirow{2}{*}{0.04} & \multirow{2}{*}{0.08} & \multirow{2}{*}{29} & \multirow{2}{*}{$35.78^{+1.02}_{-0.97}$} & \multirow{2}{*}{$35.78^{+1.02}_{-0.97}$} & 60.56(1) & 8.51(1) & 3.56 & 2.42 & 0.1163 \\
& & & & & & & 67.84(4) & 12.29(18) & 2.76 & 2.71 & 0.1249 \\
\hline
\multirow{2}{*}{QPSOe} & \multirow{2}{*}{254279323.706} & \multirow{2}{*}{0.19} & \multirow{2}{*}{0.29} & \multirow{2}{*}{18} & \multirow{2}{*}{$34.74^{+1.74}_{-1.51}$} & \multirow{2}{*}{$41.90^{+1.28}_{-1.14}$} & 37.13(1) & 4.86(3) & 3.82 & 3.71 & 0.1258 \\
 & & & & & & & 35.50(3) & 7.42(12) & 2.39 & 3.55 & 0.0026 \\
\enddata

\tablecomments{ $^{\dagger}$ The QPSO candidate persists throughout the entire burst *$E_{\rm piv}$ represents the pivot energy of the hardness ratio.}
\label{tab:qpsos}
\end{deluxetable}

%%%%%%%%%%%%%%%%%%%%%%%%%%%%%%%%%%%%%%%%%%%%%%%%%
%%%%%%%%%%%%%%%%%%%%%%%%%%%%%%%%%%%%%%%%%%%%%%%%%
%%%%%%%%%%%%%%%%%%%%%%%%%%%%%%%%%%%%%%%%%%%%%%%%%
%%%%%%%%%%%%%%%%%%%%%%%%%%%%%%%%%%%%%%%%%%%%%%%%%
%EVENT MET : 254366383 Pivot: 29 keV
%EVENT MET : 254299756 Pivot: 15 keV
%%%%%%%%%%%%%%%%%%%%%%%%%%%%%%%%%%%%%%%%%%%%%%%%%
%%%%%%%%%%%%%%%%%%%%%%%%%%%%%%%%%%%%%%%%%%%%%%%%%
%%%%%%%%%%%%%%%%%%%%%%%%%%%%%%%%%%%%%%%%%%%%%%%%%
%%%%%%%%%%%%%%%%%%%%%%%%%%%%%%%%%%%%%%%%%%%%%%%%%

The most significant QPSO candidate is in Burst MET 254366383.448 (hereafter QPSOa, \autoref{qpo_plot1} top panel). The search yields a pivot energy of 29 keV, exhibits oscillations at 29.0 Hz with a FWHM of 7.1 Hz (using WWZ) and 27.9 Hz with a FWHM of 10.9 Hz (using LSP). The corresponding $p$-values of these detections are 0.0568 and 0.0066 respectively. The QPSO spans the entire burst duration of 0.12 s. The spectral analysis revealed that the burst comprises three non-overlapping time segments, with $E_{\rm peak}$ values ranging from about 35 to 30 keV.

The second QPSO candidate which is in Burst MET 254299756.841 (QPSOb, \autoref{qpo_plot1} bottom panel) originates from the brightest unsaturated burst. This burst consists of 9 non-overlapping spectral segments and the QPSO candidate spans the burst from the beginning till the end, excluding the last segment which primarily consists of the dimmer, low-energy tail of the burst. Although we initially applied the same procedure described above to search for QPSO candidates in this burst, we detected an oscillation peak at a frequency of approximately 15 Hz, corresponding to a region dominated by red noise \citep{Huppenkothen_2013}. Thus, our previously utilized significance estimation methods (e.g., Baluev's method), which assume Gaussian white noise, are not suitable in this scenario. As an alternative, we calculated the wavelet transform and its associated significance level using the algorithm by \cite{Torrence98}, which is specifically designed to accommodate red noise assumptions. This approach is uniquely applicable to this particular burst due to its highly regular sampling, namely 91.11$\%$ of time intervals matched the minimal unbinned sampling interval of 0.002 s. Using this wavelet method, we identified oscillations with a frequency of 15.73 Hz and a corresponding FWHM of 2.67 Hz. The calculated $p$-value for this detection is 0.0001, which corresponds to a significance of $\sim3.67\sigma$. Although we could not compute a similar significance level for the Lomb-Scargle Periodogram (LSP) method due to red noise limitations, the QPSO candidate from the LSP analysis exhibits distinct harmonic-like structures around 30 Hz and 120 Hz. It is important to note that this burst stands out from other QPSO candidates in two notable ways: it exhibits a significantly lower peak frequency and has the longest duration of 0.29 s. Moreover, being the brightest unsaturated burst analyzed, it unsurprisingly displays the highest Q value of 5.89 measured in the wavelet method. Additionally, its $E_{\rm peak}$ values drop from about 20 keV at the start to 13 keV at the end of the QPSO are considerably lower than those of other candidates, which typically range from 30 to 40 keV.

The QPSO candidate in Burst MET 254291434.732 (QPSOc) is another persistent QPSO spanning the entire burst duration. This burst is relatively short, with a duration of 0.15 seconds, and consists of only one non-overlapping time segment with $E_{\rm peak}$ of 40.3 keV. The QPSO candidate exhibits oscillation frequencies of 27.8 Hz (WWZ) and 29.5 Hz (LSP), with FWHM values of 8.0 Hz and 8.1 Hz, respectively. 

The last two QPSO candidates are not persistent throughout the entire duration of their respective bursts. The QPSO candidate in burst MET 254302590.741 (QPSOd) has a broader peak with peak frequency and FWHM of 60.6 Hz with an FWHM of 17.02 Hz (WWZ) and 67.8 Hz with an FWHM of 24.6 Hz (LSP). The QPSO duration is only 0.04 s and it is considerably shorter compared to other candidates whose durations are on the order of 0.10 s. Therefore, it has a slightly higher $p$-value of around 0.12. On the other hand, the candidate in burst MET 254279323.706 (QPSOe) has peak and FWHM values of 37.13 Hz and 9.71 Hz (using WWZ) and 35.5 Hz and 14.8 Hz (using LSP). Finally, their corresponding $p$-values are 0.1258 and 0.0026, respectively.

\section{Discussion} \label{sec:dis}

\sgra burst spectra were investigated both in a time-integrated and time-resolved manner in the past. \citet{Younes14} performed time-resolved spectral analysis of 63 bursts of \sgranosp. Note that 40 out of our sample of 44 events are also included in their sample of 63 bursts. 
However, there are key methodological differences between our study and theirs. \citet{Younes14} extracted spectra using much finer time resolution, with time bins defined based on the signal-to-noise ratio. In contrast, our study segments bursts based on spectral similarity with clustering to identify natural change points in spectral evolution. Let us first compare our results to theirs. Using the COMPT model, they found a Gaussian distribution for the photon index with a mean of $-$0.55, which is compatible with our result. In addition, they found broken power law relationships between flux and the two parameters of the COMPT model with a break around $f \sim 10^{-5}$\,erg~cm$^{-2}$ s$^{-1}$. 

\begin{figure}[!htbp]
    \centering
    \epsscale{2.5}
    \includegraphics[width = \linewidth, trim = {50 90 50 0}, clip]{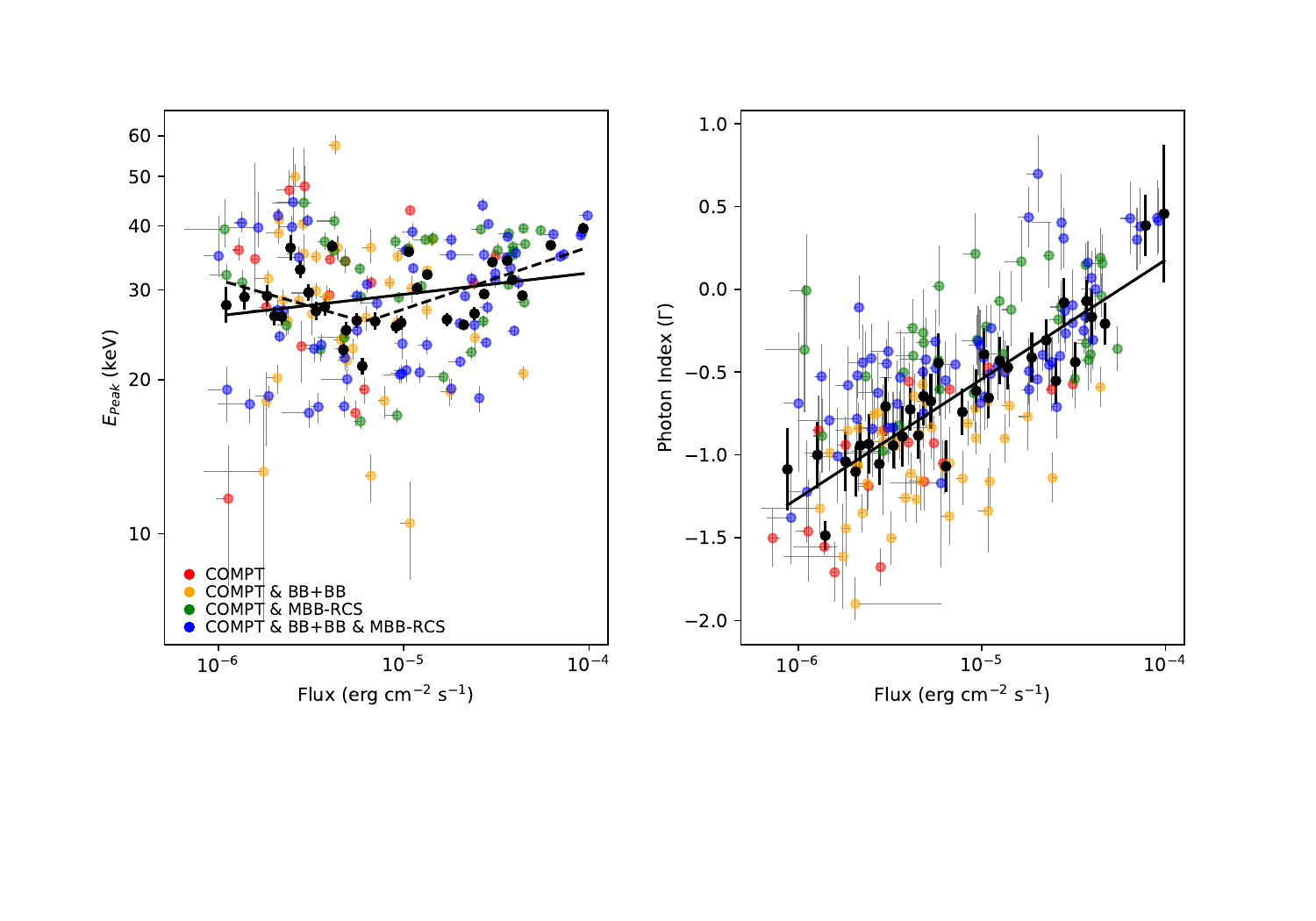}
       \caption{The scatter plot of COMPT $E_{\rm peak}$ vs. flux (left panel) and photon index vs. flux (right panel). Color code shows the preferred photon model(s) based on BIC values. The black dots represent the weighted means of consecutive groups, each with 10 data points. The black lines show the best single power law fit to the relation between the weighted means of $E_{\rm peak}$ and flux, and between the weighted means of photon index and flux, respectively. The dashed line shows the best broken power law fit between the weighted means of $E_{\rm peak}$ and flux.}

    \label{epeak_index_flux}
\end{figure}

We find a similar broken power law relation between flux and $E_{\rm peak}$ with slopes -0.10 $\pm$ 0.07, 0.12 $\pm$ 0.04 and a break point at (6.17 $\pm$ 2.41) $\times 10^{-6}$ erg cm$^{-2}$ s$^{-1}$. These values are consistent with their results within the uncertainties. However, we also find that this relation can be modeled by a single power law with a comparable BIC value ($\Delta$BIC $\approx 1$ between the single and broken power law models) with a power law index of 0.04 $\pm$ 0.02 (left panel of \autoref{epeak_index_flux}). This is likely due to the limited sample size and narrower flux range compared to the dataset of \citet{Younes14}. To further test whether our current dataset is sufficient to robustly detect a break if one truly exists, we performed a jackknife resampling analysis \citep{Tukey58}. We find that the estimated break point in the $E_{\rm peak}$–flux relation is highly stable, with a standard deviation of only 0.03 dex, which suggests the break point location is not overly sensitive to individual data points. As for the photon index vs. flux plot, they reported no correlation up to the break point of $\sim 10^{-5}$ erg cm$^{-2}$ s$^{-1}$, then a similar positive correlation as in $E_{\rm peak}$ vs. flux. We find that the flux-photon index relation can also be described by a single power law as well as a broken power law. However, the jackknife analysis reveals that the break point estimate is unstable with a standard deviation of 0.21 dex. Moreover, when employing a broken power law, only two data points remain on the high-flux side of the break point. Therefore, we adopt a single power law model with a slope of 0.72 $\pm$ 0.06 for the entire flux range (see right panel of \autoref{epeak_index_flux}).

%%%%%%%%%%%%%%%%%%%%%%%%%%%%%%%%%%%%%%%%%%%%%%%%%
%%%%%%%%%%%%%%%%%%%%%%%%%%%%%%%%%%%%%%%%%%%%%%%%%
%%%%%%%%%%%%%%%%%%%%%%%%%%%%%%%%%%%%%%%%%%%%%%%%%
%%%%%%%%%%%%%%%%%%%%%%%%%%%%%%%%%%%%%%%%%%%%%%%%%

%%%%%%%%%%%%%%%%%%%%%%%%%%%%%%%%%%%%%%%%%%%%%%%%%
%%%%%%%%%%%%%%%%%%%%%%%%%%%%%%%%%%%%%%%%%%%%%%%%%
%%%%%%%%%%%%%%%%%%%%%%%%%%%%%%%%%%%%%%%%%%%%%%%%%
%%%%%%%%%%%%%%%%%%%%%%%%%%%%%%%%%%%%%%%%%%%%%%%%%

\begin{figure}[!htbp]
    \centering
    \epsscale{1.15}
    \includegraphics[width = \linewidth,  trim = {40 100 40 20}, clip]{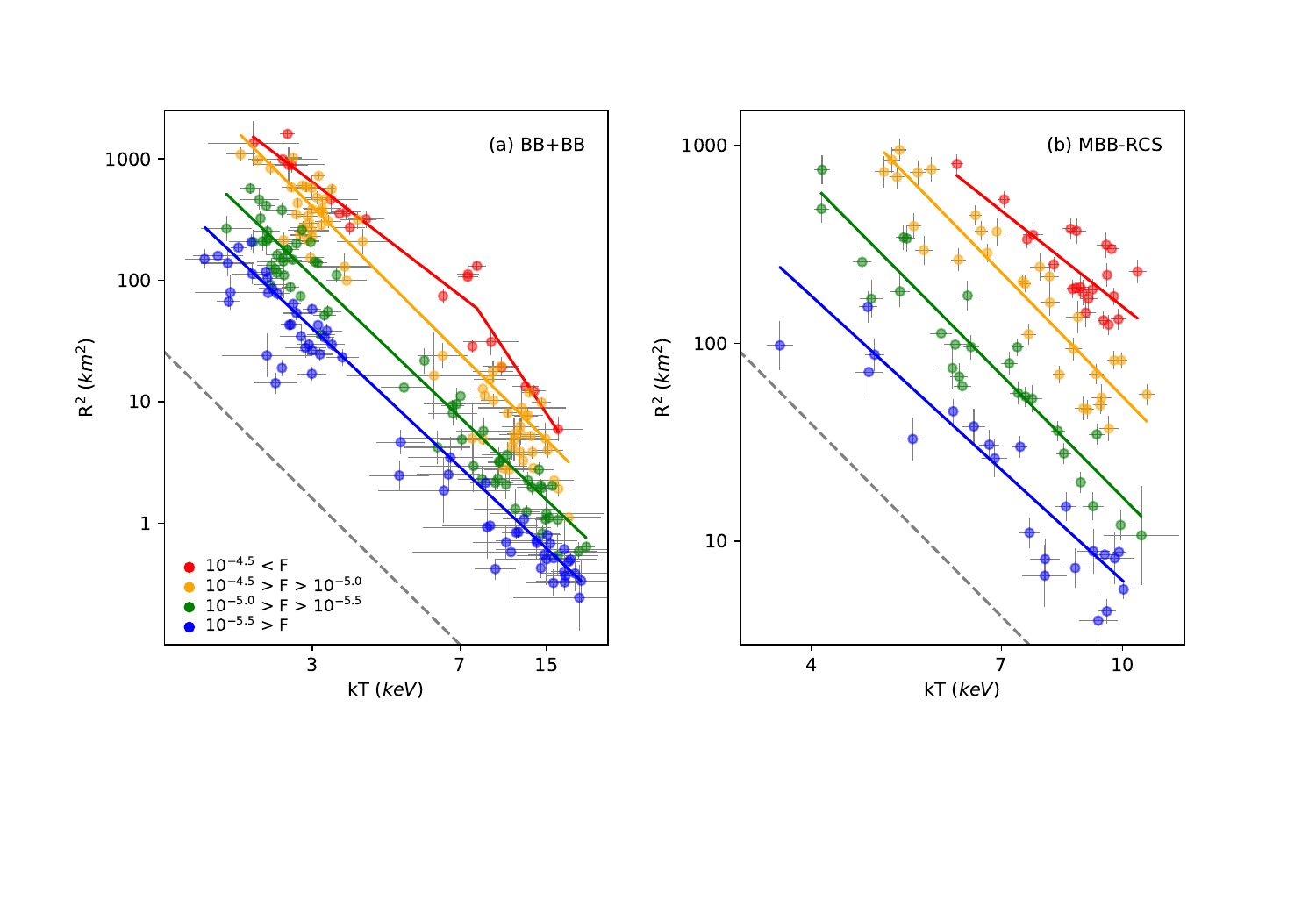}
       \caption{(left panel) Flux color-coded plot of $R^{2}$ vs. $kT$ for BB+BB. Each data point represents the weighted means of $R^{2}$ and $kT$ of every two time segments only for display purposes. Solid lines show the best-fit models. (right panel) Flux color-coded scatter plot of $R^{2}$ vs. $kT$ for MBB-RCS. Solid lines represent PL fits. We take the source distance = 5 kpc. In both panels, the gray dashed line represents $ R^2 \propto T^{-4}$.\\
}
    \label{r2_kT}
\end{figure}

\citet{Younes14} also investigated the relationship between the blackbody temperature ($kT$) and the surface area of the inferred emission region ($ R^2$) in four flux regimes. They showed that the $ R^2$ vs.~$kT$ values follow a broken power-law trend for the two parameters above $10^{-4.5}$ erg cm$^{-2}$ s$^{-1}$; for flux below this, the trend is consistent with a single power law. We find similar broken power law trend at the highest flux levels (F $> 10 ^{-4.5}$ erg cm$^{-2}$ s$^{-1}$) and power law trend can describe $ R^2$ vs.~$kT$ trends in all lower flux regimes (see \autoref{r2_kT}).

We also investigated the $ R^2$ vs.~$ kT$ behavior using the parameters of the MBB-RCS model with the same flux intervals (see the right panel of \autoref{r2_kT}). The MBB-RCS model was not used in \citet{Younes14}. We find that they are all well represented by single power laws (see \autoref{area_kT} for power-law indices). We find that in all flux regimes, the trends are consistent with the expectations of Stefan-Boltzmann law ($ R^2$ $\propto$ $ kT^{-4}$) except for a slight deviation in the highest flux regime.

We compared the results of the \sgra investigations with those of \citet{Keskin24}, who employed the same spectral analysis approach for the bursts of \sgrbnosp. In the analysis of the COMPT model, there are notable differences in the flux relationships of both parameters. In \sgrbnosp, both parameters and flux values have a broken power law relation. The correlation between $E_{\rm peak}$ and flux is positive which becomes steeper after a break value. Similarly, the photon index initially shows a positive correlation with flux but transitions to a negative correlation beyond the break. Conversely, in \sgranosp, the flux-$E_{\rm peak}$ relationship can be explained with either a slight positive correlation or a broken power law relation as described above. Thus, the flux-photon index has a strong positive correlation, with $\alpha$ = 0.72 $\pm$ 0.06.

In the analysis of the BB+BB model, there is a distinct positive correlation between $kT_{l}$ and $kT_{h}$ in both sources. Secondly, flux and both $kT$ values have a positive correlation in \sgrbnosp. In contrast, the highest flux values are accumulated around the peak of the Gaussian distribution in both $kT_{l}$ and $ kT_{ h}$ parameters of \sgranosp. Lastly, the relationship between the values of $ R^2$ and $ kT$ of BB + BB can be fitted with a broken power law in \sgrbnosp, except for the lowest flux group, which can be fitted with a single and a broken power law. On the other hand, \sgra parameters show a single power law fit characteristics with only the exception of the highest flux group that shows a broken power law relation. 

MBB-RCS model fit results show similar characteristics for the bursts of the two magnetars. Namely, $ R^2$ vs. $ kT_m$ relations exhibit single power-laws in all flux groups of both sources. It is worth mentioning that there is a significant deviation from the Stephan-Boltzmann law at the highest flux values in \sgrbnosp. This deviation is expected since the MBB-RCS model \citep{Yamasaki20} assumes photons emitted from the fireball to scatter once by the magnetospheric particles in the resonant layer. This generates a tail in the spectrum at higher energies and therefore a deviation from the Stephan-Boltzmann law. However, even at the highest flux levels ($F>10^{-4.5}$) in \sgranosp, the slope value remains close to the expected value of --4 with $\alpha=-3.12 \pm 0.5$. This suggests that a single scattering case is sufficient to account for the observed burst spectra of \sgranosp, in contrast to the deviations seen in \sgrbnosp.

In the framework proposed by the MBB-RCS model fit results for \sgrb  \citep{Keskin24}, such deviations from the  Stefan-Boltzman law were interpreted as signatures of anisotropic radiation fields near the apex of an extended magnetic flux tube. In this picture, the highly optically thick plasma in the emission region cools adiabatically as it flows from smaller, hotter regions near the flux tube footpoints on the stellar surface to the larger, cooler regions near the equatorial apex. This geometry results in the observed spectral extension. If the deviation from the canonical $R^2$–$kT^4$ relation seen in \sgrb indeed reflects the spatial extent and geometry of the active emission region—modeled as a broad, flaring flux tube—then the absence of such a deviation in \sgra may instead reflect differences in tube structure or the strength of radiative anisotropy.

Specifically, the tube length can be estimated as $R_l \sim 4R^2 / R_t$, where $R_t$ is the transverse cross-section of the flux tube at its apex. Although the observed values $R^2$ in \sgra are higher (up to $\sim$ 1000 km$^2$) than those in \sgrb (a few hundred km$^2$, \citealt{Keskin24}), this may correspond to a proportionally larger cross-section of the apex $R_t$, resulting in a thicker, more compact flux tube. In such a geometry, the anisotropy of the emergent radiation—arising from direction-dependent Compton scattering in low magnetic field regions—is likely reduced. Consequently, the observed radiation remains more isotropic, preserving the expected  Stefan-Boltzman slope even at high altitudes where anisotropic effects would otherwise dominate.

%%%%%%%%%%%%%%%%%%%%%%%%%%%%%%%%%%%%%%%%%%%%%%%%%
%%%%%%%%%%%%%%%%%%%%%%%%%%%%%%%%%%%%%%%%%%%%%%%%%
%%%%%%%%%%%%%%%%%%%%%%%%%%%%%%%%%%%%%%%%%%%%%%%%%
%%%%%%%%%%%%%%%%%%%%%%%%%%%%%%%%%%%%%%%%%%%%%%%%%

%%%%%%%%%%%%%%%%%%%%%%%%%%%%%%%%%%%%%%%%%%%%%%%%%
%%%%%%%%%%%%%%%%%%%%%%%%%%%%%%%%%%%%%%%%%%%%%%%%%
%%%%%%%%%%%%%%%%%%%%%%%%%%%%%%%%%%%%%%%%%%%%%%%%%
%%%%%%%%%%%%%%%%%%%%%%%%%%%%%%%%%%%%%%%%%%%%%%%%%

\begin{table}[!htbp]
\centering
\tabletypesize{\footnotesize}
\caption{Fit results of the trends between the blackbody emitting area ($R^2$) vs. temperature ($kT$) from MBB-RCS and BB+BB models}
    \begin{tabular}{c|c|ccc}

        & MBB-RCS & &BB+BB &  \\
    
        \hline
        \hline
        Flux Range & $ \alpha$-$ kT_{ m}$ & $ \alpha$-$ kT_{ l}$ & $ \alpha$-$ kT_{ h}$ &  kT$_{break}$ \\

        (erg cm$^{-2}$ s$^{-1}$)& & & &(keV) \\  
        \hline
        $F > 10^{-4.5}$ & $-3.12 \pm 0.5$ & $-3.10 \pm 0.30$ & $-6.12 \pm 1.26 $ & $10.72 \pm 1.12$ \\
        \hline
        $ 10^{-5.0} < F < 10^{-4.5}$ & $-4.04 \pm 0.30$ & $-4.03 \pm 0.11 ^{a}$ & $--$ & $--$ \\
        \hline
        $ 10^{-5.5} < F < 10^{-5.0}$ & $-3.99 \pm 0.30$ & $-3.86 \pm 0.08 ^{a}$ & $--$ & $--$ \\
        \hline
        $ F < 10^{-5.5}$ & $-3.61 \pm 0.31$ & $-3.80 \pm 0.06 ^{a}$ & $--$ & $--$ \\
        \hline
    
    \end{tabular}
    \tablecomments{ $^{a}$ A single PL fit to the data.
    }
\label{area_kT}
\end{table}

%%%%%%%%%%%%%%%%%%%%%%%%%%%%%%%%%%%%%%%%%%%%%%%%%
%%%%%%%%%%%%%%%%%%%%%%%%%%%%%%%%%%%%%%%%%%%%%%%%%
%%%%%%%%%%%%%%%%%%%%%%%%%%%%%%%%%%%%%%%%%%%%%%%%%
%%%%%%%%%%%%%%%%%%%%%%%%%%%%%%%%%%%%%%%%%%%%%%%%%

As far as the oscillation characteristics of QPSO candidates obtained (see \autoref{tab:qpsos}), QPSOa and QPSOc (possibly QPSOe as well) exhibit very similar oscillation frequencies of around 28 Hz. Therefore, there are two (or possibly three) bursts with a QPSO frequency of near 30 Hz, one very bright and long burst with a QPSO frequency of around 15 Hz, and another burst with a shorter QPSO at approximately 60 Hz. These frequencies hint at a potential harmonic pattern, suggesting that 28 Hz may lie near the fundamental frequency, making the 15 Hz signal its sub-harmonic and 60 Hz is the first harmonic.

Regarding the evolution of $E_{\rm peak}$ during the QPSO candidates, all except one (QPSOb) have $E_{\rm peak}$ values between 30-40 keV, which coincide with the mean and the denser central region of the parameter distribution (see \autoref{copl_combined}). No $E_{\rm peak}$ evolution is observed in QPSOc and QPSOd since both occur within a single time segment. In QPSOa, the parameter decreases from approximately 35 keV to 30 keV, whereas in QPSOe, it increases from 35 keV to around 42 keV. However, since each of these four bursts consists of at most three time segments, drawing definitive conclusions about the temporal evolution of $E_{\rm peak}$ is difficult. Conversely, the photon index parameter stays constant within their 1$\sigma$ error range throughout the time segments. 

On the other hand, QPSOb (\autoref{qpo_plot1}b) spans over 8 of the total of 9 time segments. During this period, $E_{\rm peak}$ starts at approximately 20 keV, follows a two-peaked trend similar to the light curve, reaching 22–23 keV at both peaks before decreasing to around 13 keV. Between the beginning and the end of this two peaked behavior, there are approximately 0.17 s, which is slightly more than twice the QPSO candidate period ($\sim$0.07 s). This suggests that the observed spectral modulation follows a periodic trend with each peak aligning with a cycle of the QPSO. The slight deviation from an exact integer multiple may indicate variations in the oscillation over time.

The second parameter of the COMPT model, the photon index, shows a subtle evolution in QPSO candidates with more than one time segment (QPSOs a, b, and e). During QPSOs a and b, the photon index parameter is slightly decreasing from $0.40$ to $-0.07$, and $-0.34$ to $-1.05$, respectively. 
A similar two-peaked behavior is also present in the photon index of QPSO b, though it is less distinct compared to $E_{\rm peak}$. 

The theoretical interpretation of QPSOs likely differs from standard QPO models, which are typically attributed to torsional shear oscillations in the magnetar’s crust triggered by its twisted magnetosphere. \cite{Roberts_2023} propose a novel model for QPSOs in magnetars, attributing them to density and pressure perturbations within a highly magnetized flux tube. According to this model, Thomson scattering should be dominant since the burst emission zone is highly optically thick due to high densities of radiating charges. Consequently, since Alfvén waves would be absorbed and dispersed in the dense plasma, it is unlikely that they are the cause of the observed $E_{\rm peak}$ oscillations. The proposed mechanism involves variations in density and pressure moving along a magnetized flux tube. These fluctuations cause adiabatic compression and expansion, which leads to variations in $E_{\rm peak}$.

We applied the prescription explained in \cite{Roberts_2023} as follows: $E_{\rm peak}$ value of QPSO candidates are measured in the range of 30--40 keV (with an exception of QPSOb, \autoref{tab:qpsos}), which implies a plasma temperature of approximately  $kT \sim 10$ keV. This corresponds to $\Theta = kT/m_e c^2 \approx 0.02$, which leads to a sound speed that aligns with the classical non-relativistic estimate of $c_s^2/c^2 \approx 5 \Theta/3$ (assuming an adiabatic index $\gamma = 5/3$). Therefore, the 3D sound speed is calculated as $c_s \approx 0.176c$. Finally, the time required for the pair plasma to propagate over the neutron star is $R_{NS}/c_s \sim 0.18$ ms as presented in \citet{Roberts_2023}. Assuming $R_{NS} = 10^6$ cm, a fluctuation of 35 ms ($\approx$ 28 Hz; QPSOs a, c, e) corresponds to acoustic wave propagation along a flux tube with an approximate length of $\mathcal{S} \sim 190 R_{NS}$. If the fluctuation is around 16 ms ($\approx$ 60 Hz, QPSOd), the length of the flux tube becomes $\mathcal{S} \sim 90 R_{NS}$. However, these lengths could be shorter due to adiabatic cooling along the flux tube. For the QPSOb, $E_{\rm peak}$ evolves approximately from 20 keV to 13 keV, which corresponds to $kT \sim 5$ keV. After repeating all calculations for this $kT$ value, for a fluctuation of around 67 ms ($\approx$ 15 Hz) we obtain a length of flux tube of $\mathcal{S} \sim 260 R_{NS}$. In this scheme, $\mathcal{S}$ ranges from about 100 to 250 $R_{NS}$ with our results. It is important to note that QSPOs is a new phenomenon and the sample investigated, as well as oscillations found here are indicative but limited to conclude any universal behavior.

\section*{acknowledgments}
We thank the anonymous reviewer for valuable comments and recommendations, which improved the quality and clarity of our manuscript. M.D., E.G., Y.K., and Ö.K. acknowledge support from the Scientific and Technological Research Council of Turkey (TÜB\.{I}TAK) through grant number 121F266. SY acknowledges the support from NSTC through grant numbers 113-2112-M-005-007-MY3 and 113-2811-M-005-006-.

\appendix

\section{Tables} \label{appA}
\setlength{\tabcolsep}{3pt}        % tighten columns
%\renewcommand{\arraystretch}{0.9}  % tighten rows
%\begin{table}[h]
%\centering
%\tabletypesize{\tiny}
\begin{longtable}{ccccccc}

\caption{Table of \sgra bursts used in the study.} \label{burst_list}\\
Burst Date & UTC & MET$^{a}$ & Duration$^{b}$ & No of Overlapping & No of Non-overlapping & QPSO\\
(yymmdd) &  & (s) & (s) & Time Segments & Time Segments & candidates \\ \hline
090122 & 00:56:35.317 & 254278597.317$^{*}$ & 0.517 & 3  & 1 & -  \\
090122 & 00:57:20.410 & 254278642.410       & 0.656 & 7  & 2 & -   \\
090122 & 00:58:00.842 & 254278682.842$^{*}$ & 0.524 & 2  & - & -   \\
090122 & 01:08:41.706 & 254279323.706       & 0.915 & 10 & 3 & QPSOe   \\
090122 & 01:14:45.985 & 254279687.985       & 0.300 & 8  & 2 & -   \\
090122 & 01:16:28.686 & 254279790.686$^{*}$ & 0.280 & 4  & 1 & -   \\
090122 & 01:18:10.302 & 254279892.302$^{*}$ & 0.336 & 3  & 1 & -   \\
090122 & 01:18:42.351 & 254279924.351$^{*}$$^{\dagger}$ & 0.248 & 1  & - & -   \\
090122 & 01:25:18.640 & 254280320.640       & 0.540 & 5  & 2 & -   \\
090122 & 01:28:59.988 & 254280541.988$^{\dagger}$       & 0.516 & 4  & 2 & -   \\
090122 & 02:32:53.944 & 254284375.944       & 0.296 & 4  & 2 & -   \\
090122 & 02:54:00.999 & 254285642.999$^{*}$$^{\dagger}$ & 0.698 & 1  & - & -   \\
090122 & 04:09:08.677 & 254290150.677       & 0.808 & 7  & 2 & -   \\
090122 & 04:12:33.001 & 254290355.001       & 0.849 & 6  & 2 & -   \\
090122 & 04:30:32.732 & 254291434.732$^{*}$ & 0.433 & 4  & 1 & QPSOc   \\
090122 & 04:32:19.025 & 254291541.025$^{*}$ & 0.814 & 3  & 1 & -   \\
090122 & 04:32:49.462 & 254291571.462$^{\dagger}$       & 0.687 & 12 & 5 & -   \\
090122 & 04:33:28.047 & 254291610.047$^{*}$ & 0.192 & 2  & - & -   \\
090122 & 04:34:09.362 & 254291651.362$^{\dagger}$       & 0.954 & 19 & 6 & -   \\
090122 & 04:34:18.108 & 254291660.108$^{*}$ & 0.484 & 3  & 1 & -   \\
090122 & 04:34:20.687 & 254291662.687       & 1.187 & 10 & 2 & -   \\
090122 & 04:40:06.444 & 254292008.444$^{*}$ & 0.491 & 2  & - & -   \\
090122 & 04:40:28.785 & 254292030.785$^{*}$ & 0.464 & 1  & - & -   \\
090122 & 05:14:03.372 & 254294045.372$^{\dagger}$       & 1.048 & 11 & 5 & -   \\
090122 & 05:14:29.229 & 254294071.229       & 0.891 & 10 & 3 & -   \\
090122 & 05:16:06.849 & 254294168.849       & 0.153 & 9 & 3 & -   \\
090122 & 05:16:18.303 & 254294180.303$^{*}$ & 0.616 & 2  & - & -   \\
090122 & 05:16:44.241 & 254294206.241$^{*}$ & 0.215 & 5  & 1 & -   \\
090122 & 05:52:15.165 & 254296337.165$^{*}$$^{\dagger}$ & 0.560 & 2  & - & -   \\
090122 & 06:03:35.989 & 254297017.989       & 0.448 & 14 & 4 & -   \\
090122 & 06:49:08.655 & 254299750.655       & 0.751 & 18 & 4 & -   \\
090122 & 06:49:14.841 & 254299756.841       & 1.364 & 26 & 8 & QPSOb   \\
090122 & 06:49:32.952 & 254299774.952       & 0.548 & 10 & 3 & -   \\
090122 & 06:49:44.192 & 254299786.192       & 0.609 & 8  & 2 & -   \\
090122 & 06:49:48.471 & 254299790.471$^{\dagger}$       & 0.939 & 8  & 8 & -   \\
090122 & 06:50:08.622 & 254299810.622$^{\dagger}$       & 0.390 & 12 & 3 & -   \\
090122 & 06:50:12.076 & 254299814.076       & 0.419 & 7  & 2 & -   \\
090122 & 06:50:14.271 & 254299816.271$^{*}$ & 0.281 & 6  & 1 & -   \\
090122 & 06:50:22.533 & 254299824.533$^{*}$ & 1.291 & 3  & 1 & -   \\
090122 & 06:50:49.339 & 254299851.339$^{\dagger}$       & 0.622 & 3  & 2 & -   \\
090122 & 06:50:50.911 & 254299852.911$^{*}$ & 1.844 & 2  & - & -   \\
090122 & 06:50:57.268 & 254299859.268$^{*}$$^{\dagger}$ & 0.405 & 2  & - & -   \\
090122 & 06:51:08.154 & 254299870.154$^{*}$ & 0.601 & 2  & - & -   \\
090122 & 06:51:14.791 & 254299876.791$^{\dagger}$       & 1.165 & 31 & 9 & -   \\
090122 & 06:52:00.167 & 254299922.167$^{\dagger}$       & 0.264 & 6  & 3 & -   \\
090122 & 06:52:03.979 & 254299925.979       & 0.678 & 12 & 4 & -   \\
090122 & 06:52:20.032 & 254299942.032$^{*}$ & 0.577 & 3  & 1 & -   \\
090122 & 06:59:35.546 & 254300377.546$^{\dagger}$       & 1.080 & 6  & 2 & -   \\
090122 & 07:00:58.715 & 254300460.715$^{\dagger}$       & 0.665 & 11 & 4 & -   \\
090122 & 07:26:29.656 & 254301991.656       & 0.917 & 10 & 2 & -   \\
090122 & 07:31:14.748 & 254302276.748$^{\dagger}$       & 1.420 & 21 & 5 & -   \\
090122 & 07:36:28.741 & 254302590.741$^{*}$ & 0.464 & 3  & 1 & QPSOd   \\
090122 & 07:40:15.939 & 254302817.939$^{\dagger}$       & 0.602 & 9 & 3 & -   \\
090122 & 08:36:30.674 & 254306192.674$^{\dagger}$       & 1.152 & 3  & 2 & -   \\
090122 & 10:03:04.670 & 254311386.670$^{\dagger}$       & 0.429 & 6  & 3 & -   \\
090122 & 10:16:40.374 & 254312202.374$^{*}$ & 0.387 & 4  & 1 & -   \\
090122 & 12:00:48.740 & 254318450.740$^{*}$ & 0.565 & 6  & 1 & -   \\
090122 & 15:02:15.402 & 254329337.402$^{*}$$^{\dagger}$ & 0.236 & 2  & - & -   \\
090122 & 15:35:53.655 & 254331355.655$^{\dagger}$       & 0.208 & 5  & 2 & -   \\
090122 & 23:14:54.053 & 254358896.053$^{\dagger}$       & 0.520 & 16 & 6 & -   \\
090122 & 23:31:19.876 & 254359881.876       & 0.428 & 4  & 2 & -   \\
090123 & 01:19:42.448 & 254366383.448       & 0.144 & 8  & 3 & QPSOa   \\
090123 & 02:42:10.695 & 254371330.695$^{\dagger}$       & 0.592 & 17 & 5 & -   \\
090123 & 16:54:38.064 & 254422479.064       & 0.220 & 6  & 3 & -   \\
090125 & 12:55:21.265 & 254580923.265$^{*}$ & 0.072 & 1  & - & -   \\
090125 & 23:00:36.087 & 254617238.087$^{\dagger}$       & 0.980 & 9 & 3 & -   \\
090203 & 20:00:39.494 & 255384041.494       & 0.272 & 16 & 5 & -   \\
090204 & 20:27:20.796 & 255472042.796$^{\dagger}$       & 0.236 & 5  & 3 & -   \\
090214 & 19:31:45.276 & 256332707.276$^{*}$$^{\dagger}$ & 0.188 & 1  & - & -   \\
090221 & 15:27:34.404 & 256922856.404$^{*}$$^{\dagger}$ & 0.228 & 2  & - & -   \\
090222 & 21:44:49.908 & 257031891.908$^{*}$$^{\dagger}$ & 0.108 & 1  & - & -   \\
090223 & 03:04:40.789 & 257051082.789$^{*}$$^{\dagger}$ & 0.152 & 4  & 1 & -   \\
090322 & 22:39:15.786 & 259454357.786$^{\dagger}$       & 0.592 & 9 & 3 & -   \\
090401 & 15:59:36.826 & 260294378.826       & 0.208 & 4  & 2 & -
\end{longtable}
\tablecomments{ 
$^{a}$ Mission Elapsed Time, the number of seconds since January 1, 2001 \\
$^{b}$ Duration of the Bayesian block \\
$^{*}$ Bursts with only one or none (-) non-overlapping time segments are not presented in the results. \\
$^\dagger$ Saturated burst}

%\end{table}

\bibliographystyle{aasjournal}
\bibliography{refs} 

\begin{thebibliography}{}
\expandafter\ifx\csname natexlab\endcsname\relax\def\natexlab#1{#1}\fi
\providecommand{\url}[1]{\href{#1}{#1}}
\providecommand{\dodoi}[1]{doi:~\href{http://doi.org/#1}{\nolinkurl{#1}}}
\providecommand{\doeprint}[1]{\href{http://ascl.net/#1}{\nolinkurl{http://ascl.net/#1}}}
\providecommand{\doarXiv}[1]{\href{https://arxiv.org/abs/#1}{\nolinkurl{https://arxiv.org/abs/#1}}}

\bibitem[{{Astropy Collaboration} {et~al.}(2022){Astropy Collaboration}, {Price-Whelan}, {Lim}, {Earl}, {Starkman}, {Bradley}, {Shupe}, {Patil}, {Corrales}, {Brasseur}, {N{\"o}the}, {Donath}, {Tollerud}, {Morris}, {Ginsburg}, {Vaher}, {Weaver}, {Tocknell}, {Jamieson}, {van Kerkwijk}, {Robitaille}, {Merry}, {Bachetti}, {G{\"u}nther}, {Aldcroft}, {Alvarado-Montes}, {Archibald}, {B{\'o}di}, {Bapat}, {Barentsen}, {Baz{\'a}n}, {Biswas}, {Boquien}, {Burke}, {Cara}, {Cara}, {Conroy}, {Conseil}, {Craig}, {Cross}, {Cruz}, {D'Eugenio}, {Dencheva}, {Devillepoix}, {Dietrich}, {Eigenbrot}, {Erben}, {Ferreira}, {Foreman-Mackey}, {Fox}, {Freij}, {Garg}, {Geda}, {Glattly}, {Gondhalekar}, {Gordon}, {Grant}, {Greenfield}, {Groener}, {Guest}, {Gurovich}, {Handberg}, {Hart}, {Hatfield-Dodds}, {Homeier}, {Hosseinzadeh}, {Jenness}, {Jones}, {Joseph}, {Kalmbach}, {Karamehmetoglu}, {Ka{\l}uszy{\'n}ski}, {Kelley}, {Kern}, {Kerzendorf}, {Koch}, {Kulumani}, {Lee}, {Ly}, {Ma}, {MacBride}, {Maljaars}, {Muna}, {Murphy}, {Norman},
  {O'Steen}, {Oman}, {Pacifici}, {Pascual}, {Pascual-Granado}, {Patil}, {Perren}, {Pickering}, {Rastogi}, {Roulston}, {Ryan}, {Rykoff}, {Sabater}, {Sakurikar}, {Salgado}, {Sanghi}, {Saunders}, {Savchenko}, {Schwardt}, {Seifert-Eckert}, {Shih}, {Jain}, {Shukla}, {Sick}, {Simpson}, {Singanamalla}, {Singer}, {Singhal}, {Sinha}, {Sip{\H{o}}cz}, {Spitler}, {Stansby}, {Streicher}, {{\v{S}}umak}, {Swinbank}, {Taranu}, {Tewary}, {Tremblay}, {de Val-Borro}, {Van Kooten}, {Vasovi{\'c}}, {Verma}, {de Miranda Cardoso}, {Williams}, {Wilson}, {Winkel}, {Wood-Vasey}, {Xue}, {Yoachim}, {Zhang}, {Zonca}, \& {Astropy Project Contributors}}]{Astropy_2022}
{Astropy Collaboration}, {Price-Whelan}, A.~M., {Lim}, P.~L., {et~al.} 2022, \apj, 935, 167, \dodoi{10.3847/1538-4357/ac7c74}

\bibitem[{{Atteia} {et~al.}(1987){Atteia}, {Boer}, {Hurley}, {Niel}, {Vedrenne}, {Fenimore}, {Klebesadel}, {Laros}, {Kuznetsov}, {Sunyaev}, {Terekhov}, {Kouveliotou}, {Cline}, {Dennis}, {Desai}, \& {Orwig}}]{Atteia87}
{Atteia}, J.~L., {Boer}, M., {Hurley}, K., {et~al.} 1987, \apjl, 320, L105, \dodoi{10.1086/184984}

\bibitem[{{Atwood} {et~al.}(2009){Atwood}, {Abdo}, {Ackermann}, {Althouse}, {Anderson}, {Axelsson}, {Baldini}, {Ballet}, {Band}, {Barbiellini}, {Bartelt}, {Bastieri}, {Baughman}, {Bechtol}, {B{\'e}d{\'e}r{\`e}de}, {Bellardi}, {Bellazzini}, {Berenji}, {Bignami}, {Bisello}, {Bissaldi}, {Blandford}, {Bloom}, {Bogart}, {Bonamente}, {Bonnell}, {Borgland}, {Bouvier}, {Bregeon}, {Brez}, {Brigida}, {Bruel}, {Burnett}, {Busetto}, {Caliandro}, {Cameron}, {Caraveo}, {Carius}, {Carlson}, {Casandjian}, {Cavazzuti}, {Ceccanti}, {Cecchi}, {Charles}, {Chekhtman}, {Cheung}, {Chiang}, {Chipaux}, {Cillis}, {Ciprini}, {Claus}, {Cohen-Tanugi}, {Condamoor}, {Conrad}, {Corbet}, {Corucci}, {Costamante}, {Cutini}, {Davis}, {Decotigny}, {DeKlotz}, {Dermer}, {de Angelis}, {Digel}, {do Couto e Silva}, {Drell}, {Dubois}, {Dumora}, {Edmonds}, {Fabiani}, {Farnier}, {Favuzzi}, {Flath}, {Fleury}, {Focke}, {Funk}, {Fusco}, {Gargano}, {Gasparrini}, {Gehrels}, {Gentit}, {Germani}, {Giebels}, {Giglietto}, {Giommi}, {Giordano}, {Glanzman},
  {Godfrey}, {Grenier}, {Grondin}, {Grove}, {Guillemot}, {Guiriec}, {Haller}, {Harding}, {Hart}, {Hays}, {Healey}, {Hirayama}, {Hjalmarsdotter}, {Horn}, {Hughes}, {J{\'o}hannesson}, {Johansson}, {Johnson}, {Johnson}, {Johnson}, {Johnson}, {Kamae}, {Katagiri}, {Kataoka}, {Kavelaars}, {Kawai}, {Kelly}, {Kerr}, {Klamra}, {Kn{\"o}dlseder}, {Kocian}, {Komin}, {Kuehn}, {Kuss}, {Landriu}, {Latronico}, {Lee}, {Lee}, {Lemoine-Goumard}, {Lionetto}, {Longo}, {Loparco}, {Lott}, {Lovellette}, {Lubrano}, {Madejski}, {Makeev}, {Marangelli}, {Massai}, {Mazziotta}, {McEnery}, {Menon}, {Meurer}, {Michelson}, {Minuti}, {Mirizzi}, {Mitthumsiri}, {Mizuno}, {Moiseev}, {Monte}, {Monzani}, {Moretti}, {Morselli}, {Moskalenko}, {Murgia}, {Nakamori}, {Nishino}, {Nolan}, {Norris}, {Nuss}, {Ohno}, {Ohsugi}, {Omodei}, {Orlando}, {Ormes}, {Paccagnella}, {Paneque}, {Panetta}, {Parent}, {Pearce}, {Pepe}, {Perazzo}, {Pesce-Rollins}, {Picozza}, {Pieri}, {Pinchera}, {Piron}, {Porter}, {Poupard}, {Rain{\`o}}, {Rando}, {Rapposelli}, {Razzano},
  {Reimer}, {Reimer}, {Reposeur}, {Reyes}, {Ritz}, {Rochester}, {Rodriguez}, {Romani}, {Roth}, {Russell}, {Ryde}, {Sabatini}, {Sadrozinski}, {Sanchez}, {Sander}, {Sapozhnikov}, {Parkinson}, {Scargle}, {Schalk}, \& {Scolieri}}]{Atwood_2009}
{Atwood}, W.~B., {Abdo}, A.~A., {Ackermann}, M., {et~al.} 2009, \apj, 697, 1071, \dodoi{10.1088/0004-637X/697/2/1071}

\bibitem[{{Baluev}(2008)}]{Baluev_2008}
{Baluev}, R.~V. 2008, \mnras, 385, 1279, \dodoi{10.1111/j.1365-2966.2008.12689.x}

\bibitem[{{Bochenek} {et~al.}(2020){Bochenek}, {Ravi}, {Belov}, {Hallinan}, {Kocz}, {Kulkarni}, \& {McKenna}}]{Bochenek_2020}
{Bochenek}, C.~D., {Ravi}, V., {Belov}, K.~V., {et~al.} 2020, \nat, 587, 59, \dodoi{10.1038/s41586-020-2872-x}

\bibitem[{Camilo {et~al.}(2007)Camilo, Ransom, Halpern, \& Reynolds}]{Camilo_2007}
Camilo, F., Ransom, S.~M., Halpern, J.~P., \& Reynolds, J. 2007, The Astrophysical Journal, 666, \dodoi{10.1086/521826}

\bibitem[{{Cash}(1979)}]{Castor}
{Cash}, W. 1979, \apj, 228, 939, \dodoi{10.1086/156922}

\bibitem[{Collazzi {et~al.}(2015)Collazzi, Kouveliotou, Horst, Younes, Kaneko, Göğüş, Lin, Granot, Finger, Chaplin, \& et~al.}]{Collazzi_2015}
Collazzi, A.~C., Kouveliotou, C., Horst, A.~J., {et~al.} 2015, The Astrophysical Journal Supplement Series, 218, 11, \dodoi{10.1088/0067-0049/218/1/11}

\bibitem[{{Duncan}(1998)}]{Duncan_98}
{Duncan}, R.~C. 1998, \apjl, 498, L45, \dodoi{10.1086/311303}

\bibitem[{{Duncan} \& {Thompson}(1992)}]{DT92}
{Duncan}, R.~C., \& {Thompson}, C. 1992, \apjl, 392, L9, \dodoi{10.1086/186413}

\bibitem[{Feroci {et~al.}(2004)Feroci, Caliandro, Massaro, Mereghetti, \& Woods}]{Feroci_2004}
Feroci, M., Caliandro, G.~A., Massaro, E., Mereghetti, S., \& Woods, P.~M. 2004, The Astrophysical Journal, 612, 408–413, \dodoi{10.1086/422405}

\bibitem[{{Foster}(1996)}]{Foster_1996}
{Foster}, G. 1996, \aj, 112, 1709, \dodoi{10.1086/118137}

\bibitem[{{Huppenkothen} {et~al.}(2013){Huppenkothen}, {Watts}, {Uttley}, {van der Horst}, {van der Klis}, {Kouveliotou}, {G{\"o}{\v{g}}{\"u}{\c{s}}}, {Granot}, {Vaughan}, \& {Finger}}]{Huppenkothen_2013}
{Huppenkothen}, D., {Watts}, A.~L., {Uttley}, P., {et~al.} 2013, \apj, 768, 87, \dodoi{10.1088/0004-637X/768/1/87}

\bibitem[{{Huppenkothen} {et~al.}(2014){Huppenkothen}, {D'Angelo}, {Watts}, {Heil}, {van der Klis}, {van der Horst}, {Kouveliotou}, {Baring}, {G{\"o}{\u{g}}{\"u}{\c{s}}}, {Granot}, {Kaneko}, {Lin}, {von Kienlin}, \& {Younes}}]{Huppenkothen_2014}
{Huppenkothen}, D., {D'Angelo}, C., {Watts}, A.~L., {et~al.} 2014, \apj, 787, 128, \dodoi{10.1088/0004-637X/787/2/128}

\bibitem[{{Hurley} {et~al.}(1999){Hurley}, {Cline}, {Mazets}, {Barthelmy}, {Butterworth}, {Marshall}, {Palmer}, {Aptekar}, {Golenetskii}, {Il'Inskii}, {Frederiks}, {McTiernan}, {Gold}, \& {Trombka}}]{Hurley99}
{Hurley}, K., {Cline}, T., {Mazets}, E., {et~al.} 1999, \nat, 397, 41, \dodoi{10.1038/16199}

\bibitem[{{Israel} {et~al.}(2005){Israel}, {Belloni}, {Stella}, {Rephaeli}, {Gruber}, {Casella}, {Dall'Osso}, {Rea}, {Persic}, \& {Rothschild}}]{Israel_2005}
{Israel}, G.~L., {Belloni}, T., {Stella}, L., {et~al.} 2005, \apjl, 628, L53, \dodoi{10.1086/432615}

\bibitem[{Israel {et~al.}(2008)Israel, Romano, Mangano, Dall’Osso, Chincarini, Stella, Campana, Belloni, Tagliaferri, Blustin, \& et~al.}]{Israel_2008}
Israel, G.~L., Romano, P., Mangano, V., {et~al.} 2008, The Astrophysical Journal, 685, 1114–1128, \dodoi{10.1086/590486}

\bibitem[{{Kaastra}(2017)}]{Kaastra2017}
{Kaastra}, J.~S. 2017, \aap, 605, A51, \dodoi{10.1051/0004-6361/201629319}

\bibitem[{{Kaneko} {et~al.}(2010){Kaneko}, {G{\"o}{\v{g}}{\"u}{\c{s}}}, {Kouveliotou}, {Granot}, {Ramirez-Ruiz}, {van der Horst}, {Watts}, {Finger}, {Gehrels}, {Pe'er}, {van der Klis}, {von Kienlin}, {Wachter}, {Wilson-Hodge}, \& {Woods}}]{Kaneko09}
{Kaneko}, Y., {G{\"o}{\v{g}}{\"u}{\c{s}}}, E., {Kouveliotou}, C., {et~al.} 2010, \apj, 710, 1335, \dodoi{10.1088/0004-637X/710/2/1335}

\bibitem[{Kass \& Raftery(1995)}]{Bayes95}
Kass, R.~E., \& Raftery, A.~E. 1995, Journal of the American Statistical Association, 90, 773, \dodoi{10.1080/01621459.1995.10476572}

\bibitem[{{Keskin} {et~al.}(2024){Keskin}, {G{\"o}{\u{g}}{\"u}{\c{s}}}, {Kaneko}, {Demirer}, {Yamasaki}, {Baring}, {Lin}, {Roberts}, \& {Kouveliotou}}]{Keskin24}
{Keskin}, {\"O}., {G{\"o}{\u{g}}{\"u}{\c{s}}}, E., {Kaneko}, Y., {et~al.} 2024, \apj, 965, 130, \dodoi{10.3847/1538-4357/ad2fce}

\bibitem[{Kırmızıbayrak {et~al.}(2017)Kırmızıbayrak, Şaşmaz Muş, Kaneko, \& Göğüş}]{Kbayrak_2017}
Kırmızıbayrak, D., Şaşmaz Muş, S., Kaneko, Y., \& Göğüş, E. 2017, The Astrophysical Journal Supplement Series, 232, 17, \dodoi{10.3847/1538-4365/aa88b7}

\bibitem[{Lamb \& Markert(1981)}]{Lamb_Markert_1981}
Lamb, R.~C., \& Markert, T.~H. 1981, The Astrophysical Journal, 244, 94, \dodoi{10.1086/158688}

\bibitem[{Laros {et~al.}(1987)Laros, Fenimore, Klebesadel, Atteia, Boer, Hurley, Niel, Vedrenne, Kane, Kouveliotou, \& et~al.}]{Laros__1987}
Laros, J.~G., Fenimore, E.~E., Klebesadel, R.~W., {et~al.} 1987, The Astrophysical Journal, 320, \dodoi{10.1086/184985}

\bibitem[{{Li} {et~al.}(2022){Li}, {Ge}, {Lin}, {Zhang}, {Song}, {Cao}, {Zhang}, {Lu}, {Xu}, {Xiong}, {Tuo}, {Tan}, {Jiang}, {Qu}, {Zhang}, {Wang}, {Wang}, {Zhang}, {Zhang}, {Li}, {Liu}, {Li}, {Bu}, {Cai}, {Chen}, {Chen}, {Chang}, {Chen}, {Chen}, {Chen}, {Cui}, {Du}, {Gao}, {Gao}, {Gu}, {Guan}, {Guo}, {Han}, {Huang}, {Huo}, {Jia}, {Jin}, {Kong}, {Li}, {Li}, {Li}, {Li}, {Li}, {Li}, {Liang}, {Liao}, {Liu}, {Liu}, {Liu}, {Lu}, {Luo}, {Luo}, {Ma}, {Ma}, {Ma}, {Meng}, {Nang}, {Nie}, {Ou}, {Ren}, {Sai}, {Song}, {Sun}, {Tao}, {Wang}, {Wang}, {Wang}, {Wang}, {Wen}, {Wu}, {Wu}, {Wu}, {Xiao}, {Yang}, {Yang}, {Yi}, {Yin}, {You}, {Yu}, {Zhang}, {Zhang}, {Zhang}, {Zhang}, {Zhang}, {Zhang}, {Zhang}, {Zhao}, {Zhao}, {Zheng}, \& {Zhou}}]{Li_2022}
{Li}, X., {Ge}, M., {Lin}, L., {et~al.} 2022, \apj, 931, 56, \dodoi{10.3847/1538-4357/ac6587}

\bibitem[{Lin {et~al.}(2011)Lin, Kouveliotou, Baring, van~der Horst, Guiriec, Woods, Göğüş, Kaneko, Scargle, Granot, \& et~al.}]{Lin_2011}
Lin, L., Kouveliotou, C., Baring, M.~G., {et~al.} 2011, The Astrophysical Journal, 739, 87, \dodoi{10.1088/0004-637x/739/2/87}

\bibitem[{{Lin} {et~al.}(2012){Lin}, {G{\"o}{\v{g}}{\"u}{\c{s}}}, {Baring}, {Granot}, {Kouveliotou}, {Kaneko}, {van der Horst}, {Gruber}, {von Kienlin}, {Younes}, {Watts}, \& {Gehrels}}]{Lin12}
{Lin}, L., {G{\"o}{\v{g}}{\"u}{\c{s}}}, E., {Baring}, M.~G., {et~al.} 2012, \apj, 756, 54, \dodoi{10.1088/0004-637X/756/1/54}

\bibitem[{{Lomb}(1976)}]{Lomb_1976}
{Lomb}, N.~R. 1976, \apss, 39, 447, \dodoi{10.1007/BF00648343}

\bibitem[{Lyubarsky(2002)}]{Lyubarsky_2002}
Lyubarsky, Y.~E. 2002, Monthly Notices of the Royal Astronomical Society, 332, 199–204, \dodoi{10.1046/j.1365-8711.2002.05290.x}

\bibitem[{{Lyutikov}(2003)}]{lyu03}
{Lyutikov}, M. 2003, \mnras, 346, 540, \dodoi{10.1046/j.1365-2966.2003.07110.x}

\bibitem[{MacQueen(1967)}]{MacQueen1967}
MacQueen, J.~B. 1967, in Proc. of the fifth Berkeley Symposium on Mathematical Statistics and Probability, ed. L.~M.~L. Cam \& J.~Neyman, Vol.~1 (University of California Press), 281--297

\bibitem[{Mazets {et~al.}(1979)Mazets, Golenetskii, Il’inskii, Aptekar’, \& Guryan}]{Mazets_1979}
Mazets, E.~P., Golenetskii, S.~V., Il’inskii, V.~N., Aptekar’, R.~L., \& Guryan, Y.~A. 1979, Nature, 282, 587–589, \dodoi{10.1038/282587a0}

\bibitem[{{Meegan} {et~al.}(2009){Meegan}, {Lichti}, {Bhat}, {Bissaldi}, {Briggs}, {Connaughton}, {Diehl}, {Fishman}, {Greiner}, {Hoover}, {van der Horst}, {von Kienlin}, {Kippen}, {Kouveliotou}, {McBreen}, {Paciesas}, {Preece}, {Steinle}, {Wallace}, {Wilson}, \& {Wilson-Hodge}}]{Meegan2009}
{Meegan}, C., {Lichti}, G., {Bhat}, P.~N., {et~al.} 2009, \apj, 702, 791, \dodoi{10.1088/0004-637X/702/1/791}

\bibitem[{{Mereghetti} {et~al.}(2009){Mereghetti}, {G{\"o}tz}, {Weidenspointner}, {von Kienlin}, {Esposito}, {Tiengo}, {Vianello}, {Israel}, {Stella}, {Turolla}, {Rea}, \& {Zane}}]{Mereghetti09}
{Mereghetti}, S., {G{\"o}tz}, D., {Weidenspointner}, G., {et~al.} 2009, \apjl, 696, L74, \dodoi{10.1088/0004-637X/696/1/L74}

\bibitem[{{Palmer} {et~al.}(2005){Palmer}, {Barthelmy}, {Gehrels}, {Kippen}, {Cayton}, {Kouveliotou}, {Eichler}, {Wijers}, {Woods}, {Granot}, {Lyubarsky}, {Ramirez-Ruiz}, {Barbier}, {Chester}, {Cummings}, {Fenimore}, {Finger}, {Gaensler}, {Hullinger}, {Krimm}, {Markwardt}, {Nousek}, {Parsons}, {Patel}, {Sakamoto}, {Sato}, {Suzuki}, \& {Tueller}}]{Palmer05}
{Palmer}, D.~M., {Barthelmy}, S., {Gehrels}, N., {et~al.} 2005, \nat, 434, 1107, \dodoi{10.1038/nature03525}

\bibitem[{Pedregosa {et~al.}(2011)Pedregosa, Varoquaux, Gramfort, Michel, Thirion, Grisel, Blondel, Prettenhofer, Weiss, Dubourg, Vanderplas, Passos, Cournapeau, Brucher, Perrot, \& {{\'E}}douard Duchesnay}]{scikit-learn}
Pedregosa, F., Varoquaux, G., Gramfort, A., {et~al.} 2011, Journal of Machine Learning Research, 12, 2825.
\newblock \url{http://jmlr.org/papers/v12/pedregosa11a.html}

\bibitem[{Roberts {et~al.}(2023)Roberts, Baring, Huppenkothen, Kouveliotou, Göğüş, Kaneko, Lin, van~der Horst, \& Younes}]{Roberts_2023}
Roberts, O.~J., Baring, M.~G., Huppenkothen, D., {et~al.} 2023, The Astrophysical Journal Letters, 956, L27, \dodoi{10.3847/2041-8213/acfcad}

\bibitem[{{Scargle}(1982)}]{Scargle_1982}
{Scargle}, J.~D. 1982, \apj, 263, 835, \dodoi{10.1086/160554}

\bibitem[{{Scargle} {et~al.}(2013){Scargle}, {Norris}, {Jackson}, \& {Chiang}}]{scargle2013}
{Scargle}, J.~D., {Norris}, J.~P., {Jackson}, B., \& {Chiang}, J. 2013, \apj, 764, 167, \dodoi{10.1088/0004-637X/764/2/167}

\bibitem[{{Schwarz}(1978)}]{Schwarz78}
{Schwarz}, G. 1978, Annals of Statistics, 6, 461

\bibitem[{{Strohmayer} \& {Watts}(2005)}]{Strohmayer_2005}
{Strohmayer}, T.~E., \& {Watts}, A.~L. 2005, \apjl, 632, L111, \dodoi{10.1086/497911}

\bibitem[{{Strohmayer} \& {Watts}(2006)}]{Strohmayer_2006}
---. 2006, \apj, 653, 593, \dodoi{10.1086/508703}

\bibitem[{Thompson \& Duncan(1995)}]{Thompson_Duncan_1995}
Thompson, C., \& Duncan, R.~C. 1995, Monthly Notices of the Royal Astronomical Society, 275, 255–300, \dodoi{10.1093/mnras/275.2.255}

\bibitem[{{Torrence} \& {Compo}(1998)}]{Torrence98}
{Torrence}, C., \& {Compo}, G.~P. 1998, Bulletin of the American Meteorological Society, 79, 61, \dodoi{10.1175/1520-0477(1998)079<0061:APGTWA>2.0.CO;2}

\bibitem[{Tukey(1958)}]{Tukey58}
Tukey, J.~W. 1958, The Annals of Mathematical Statistics, 29, 614 , \dodoi{10.1214/aoms/1177706647}

\bibitem[{{van der Horst} {et~al.}(2012){van der Horst}, {Kouveliotou}, {Gorgone}, {Kaneko}, {Baring}, {Guiriec}, {G{\"o}{\v{g}}{\"u}{\c{s}}}, {Granot}, {Watts}, {Lin}, {Bhat}, {Bissaldi}, {Chaplin}, {Finger}, {Gehrels}, {Gibby}, {Giles}, {Goldstein}, {Gruber}, {Harding}, {Kaper}, {von Kienlin}, {van der Klis}, {McBreen}, {Mcenery}, {Meegan}, {Paciesas}, {Pe'er}, {Preece}, {Ramirez-Ruiz}, {Rau}, {Wachter}, {Wilson-Hodge}, {Woods}, \& {Wijers}}]{von12}
{van der Horst}, A.~J., {Kouveliotou}, C., {Gorgone}, N.~M., {et~al.} 2012, \apj, 749, 122, \dodoi{10.1088/0004-637X/749/2/122}

\bibitem[{{van der Klis}(2006)}]{vanderklis_2006}
{van der Klis}, M. 2006, in Compact stellar X-ray sources, ed. W.~H.~G. {Lewin} \& M.~{van der Klis}, Vol.~39, 39--112

\bibitem[{{von Kienlin} {et~al.}(2012){von Kienlin}, {Gruber}, {Kouveliotou}, {Granot}, {Baring}, {G{\"o}{\v{g}}{\"u}{\c{s}}}, {Huppenkothen}, {Kaneko}, {Lin}, {Watts}, {Bhat}, {Guiriec}, {van der Horst}, {Bissaldi}, {Greiner}, {Meegan}, {Paciesas}, {Preece}, \& {Rau}}]{vonkienlin2012}
{von Kienlin}, A., {Gruber}, D., {Kouveliotou}, C., {et~al.} 2012, \apj, 755, 150, \dodoi{10.1088/0004-637X/755/2/150}

\bibitem[{{Xiao} {et~al.}(2024){Xiao}, {Li}, {Xue}, {Xiong}, {Zhang}, {Peng}, {Dong}, {Tuo}, {Cai}, {Luo}, {Yang}, {Wang}, {Zheng}, {Zhang}, {Liu}, {Tan}, {Wang}, {Wang}, {Li}, {Yi}, {Dang}, {Shang}, {Zhao}, {Ma}, {Xie}, {Feng}, {Zhang}, {Zhang}, {Ge}, {Zheng}, {Song}, \& {Zhi}}]{Xiao_2024}
{Xiao}, S., {Li}, X.-B., {Xue}, W.-C., {et~al.} 2024, \mnras, 527, 11915, \dodoi{10.1093/mnras/stae009}

\bibitem[{{Yamasaki} {et~al.}(2020){Yamasaki}, {Lyubarsky}, {Granot}, \& {G{\"o}{\u{g}}{\"u}{\c{s}}}}]{Yamasaki20}
{Yamasaki}, S., {Lyubarsky}, Y., {Granot}, J., \& {G{\"o}{\u{g}}{\"u}{\c{s}}}, E. 2020, \mnras, 498, 484, \dodoi{10.1093/mnras/staa2223}

\bibitem[{{Younes} {et~al.}(2014){Younes}, {Kouveliotou}, {van der Horst}, {Baring}, {Granot}, {Watts}, {Bhat}, {Collazzi}, {Gehrels}, {Gorgone}, {G{\"o}{\u{g}}{\"u}{\c{s}}}, {Gruber}, {Grunblatt}, {Huppenkothen}, {Kaneko}, {von Kienlin}, {van der Klis}, {Lin}, {Mcenery}, {van Putten}, \& {Wijers}}]{Younes14}
{Younes}, G., {Kouveliotou}, C., {van der Horst}, A.~J., {et~al.} 2014, \apj, 785, 52, \dodoi{10.1088/0004-637X/785/1/52}

\end{thebibliography}

\end{document}